\documentstyle[nato,epsf]{crckapb} 

\begin{opening}
\title{EXPERIMENTAL TESTS OF THE STANDARD MODEL}
\author{L. Nodulman}
\institute{Argonne National Laboratory\\
           High Energy Physics Division\\
	   Argonne IL 60439 USA}

\end{opening}

\runningtitle{STANDARD MODEL TESTS}

\begin{document}


\section{Introduction}

The title
implies an impossibly broad field, as the Standard Model includes
the fermion matter states, as well as the forces and fields of 
$SU(3) \times SU(2) \times U(1)$.  For practical purposes, I will confine
myself to electroweak unification, as discussed in the lectures of 
M. Herrero.  Quarks and mixing were discussed in the lectures of 
R. Aleksan, and leptons and mixing were discussed in the lectures of
K. Nakamura.  I will essentially assume universality, that is flavor
independence, rather than discussing tests of it.

I will not pursue tests of QED beyond noting the consistency and
precision of measurements of $\alpha_{EM}$ in various processes including
the Lamb shift, the anomalous magnetic moment (g-2) of the electron, and
the quantum Hall effect.  The fantastic precision and agreement of these
predictions and measurements is something that 
convinces people that there may be
something to this science enterprise.

Also impressive is the success of the ``Universal Fermi Interaction''
description of beta decay processes, or in more modern parlance,
weak charged current interactions.  With one coupling constant $G_F$, most
precisely determined in muon decay, a huge number of nuclear instabilities
are described.  The slightly slow rate for neutron beta decay was one
of the initial pieces of evidence for Cabbibo mixing, 
now generalized so that all charged
current decays of any flavor are covered.  

QCD has also evolved an impressive ability to predict a wide range of
measurements with a universal coupling, $\alpha_S$.  
Tests of QCD were covered in 
the lectures of J. Stirling.  Clearly the issues of associating final
state jets with quarks and gluons, and of analyzing 
proton structure in terms of quarks and
gluons will be important in many experimental tests of electroweak
unification.

The lack of renormalizability of the Fermi theory of charge current weak 
interactions, that is the inability to calculate radiative corrections, and thus
bad behavior in the high energy limit, was the motivation for models of
unification.  One of several such models, which, in the early
70's went under the name ``Weinberg-Salam,''\cite{ws} has become ``Standard.''
A simple-minded picture of this model is that by combining an isosinglet
and a isotriplet of gauge bosons, one mixes up the $\gamma$ for QED, 
heavy $W^{\pm}$
bosons for the weak charged current, and a heavy $Z^0$ which predicted
the weak neutral current interaction. A third parameter describing 
the neutral weak 
interaction can be taken as some definition of the weak neutral 
triplet/singlet mixing angle,
$\Theta_W$, or more practically, 
as the rather precisely measured mass of the $Z^0$ boson. A fourth
parameter is needed, associated with consistency in heavy gauge boson masses;
this may be taken as the Standard Model Higgs mass, which is largely decoupled
from observables.

For practical purposes, I will consider the Standard Model to include the
simplest Higgs mechanism, one complex doublet, 
with one residual Higgs particle with a mostly
unpredicted mass.  Implementation of the Higgs mechanism 
in terms of fundamental scalar multiplets 
may well be just a mathematical trick; one certainly hopes that nature
could not really be so unimaginative. But since the simplest scheme is still
viable, I will take it as standard.  I note that in terms of multiple
Higgs states, infinite variety is possible, although there are some
constraints from measurements of the $\rho$ parameter,
as M. Herrero discussed in terms of ``custodial symmetry.''

The unified electroweak theory does allow calculation of radiative
corrections. These corrections give terms involving squares of
fermion masses, and logarithms of scalar masses, for example, in
predicting the weak mixing angle and/or the W boson mass. That is, heavier
masses imply larger effects.  These would include particles we may not know
about, as well as particles regarded as standard. So if everything hangs 
together in terms of the top quark mass and the Higgs mass, we obtain 
constraints on what else could be out there.  

Precision measurements to challenge these predictions have been an essential
feature of $e^+e^-$ collider studies culminating in the LEP and SLC programs.
The predictions serve
as a motivator for high energy hadron collider programs, from the
$S\bar{p}pS$ collider at CERN, to the current Fermilab Tevatron Collider, 
and the LHC being built at CERN.
Most processes involve ``oblique'' or propagator corrections; these involve
top mass squared and Higgs mass logarithmic terms.  Notably the $Z^0$
decay rate to $b$ pairs, $R_b$, depends on vertex corrections which 
depend essentially on the top mass.  One of the claims of success of
this precision measurement program has been the consistency of the
indirectly implied top mass with the hadron collider top mass 
limits and, eventually, top mass
measurements.  Thus, the measurement of the top mass is an essential
part of the program, which has moved on to constraining the Higgs.

I will review several measurements to illustrate this program: 
The new muon (g-2) experiment at Brookhaven illustrates both weak and
QED measurement.  The NuTeV experiment at Fermilab illustrates the
historically important and currently still competitive contribution of
neutrino physics.  The LEP precision $Z^0$ lineshape and $Z^0$ decay
asymmetry measurements are clearly the core of this program. The ultimate
$Z^0$ asymmetry measurement comes, of course, from SLC. The increasingly
precise $W$ boson mass measurements at the Tevatron have been joined by
measurements at LEP2.  The top mass measurements complete the
indirect Higgs picture, but the direct Higgs search at LEP2 has
a significant impact as well.

\section{BNL E815 MUON (g-2)}

\subsection{GOALS OF THE MEASUREMENT}

The anomalous magnetic moment of the muon, defined as \\
\vspace{-.1in}
\begin{center}
$a_{\mu} \equiv \mu_{\mu}/(e\hbar/2m_{\mu})-1 = (g_{\mu}-2)/2$,\\ 
\end{center}
\vspace{.1in}
is a less
favorable QED test than the electron magnetic moment. The heavier muon
mass makes radiative corrections involving hadronic states relatively
important.  The heavier muon mass
also makes electroweak radiative corrections relatively
important.  Previous measurements at CERN\cite{Bailey} were precise
enough to establish the presence of hadronic corrections.  The goal at 
Brookhaven is to measure $a_{\mu}$ well enough to get a handle on the 
electroweak radiative corrections; a requirement for this is a precise
enough prediction of hadronic corrections to allow the EWK corrections to
be isolated.\cite{roberts} 

\begin{table}[htb]
\begin{center}
\caption{Values and corrections to $a_{\mu}$ in units of $10^{-11}$.}
\begin{tabular}{lrr}
\hline
Quantity & Value & Error\\
\hline
QED prediction & 116584706 & 2\\
EWK correction & 151 & 4\\
HAD correction & 6771 & 77\\
Overall prediction & 116591628 & 77\\
CERN measurement & 116592300 & 840\\
\hline
\end{tabular}
\end{center}
\end{table}

The magnetic moment numbers are
given in Table 1. The QED prediction is calculated
to $\alpha^5$\cite{ew2b} and the EWK correction is calculated to two 
loops.\cite{ew2b,ew2f}
The leading order hadronic correction, a vacuum polarization loop with hadrons,
must be determined with dispersion relations, using
the low energy $e^+e^-\rightarrow hadrons$
measurements.\cite{had1d} These measurements
are being improved at CMD2 in Novosibirsk, BES, and in $\tau$ decay studies.
Higher order hadronic corrections, and light-by-light or
diagrams with four photons converging on a loop\cite{lol} are relatively under
control.

\begin{figure}
\vspace{.1in}
\hspace{-.01in}\epsfysize4.3in\epsffile[40 300 560 750]{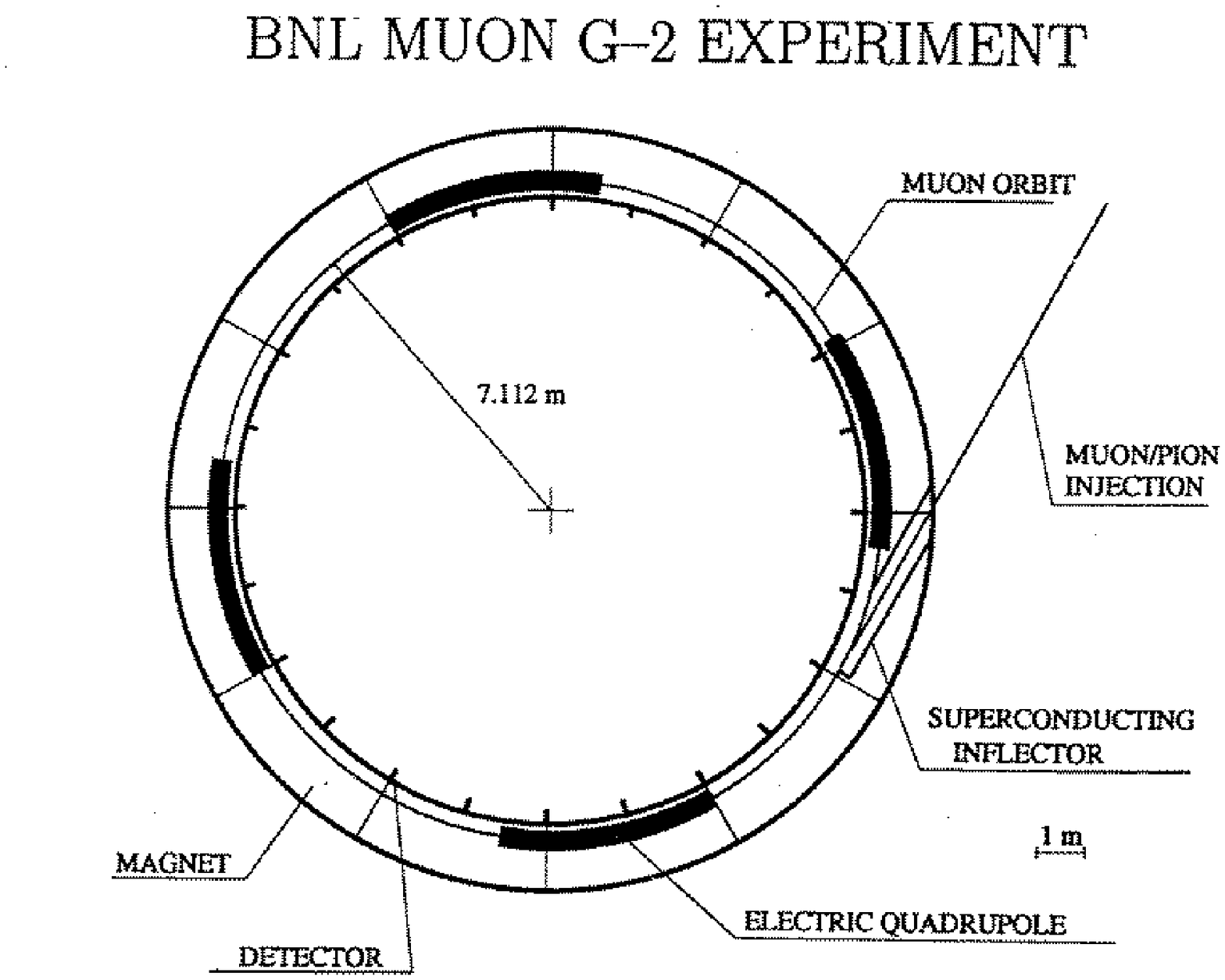}
\caption{The Brookhaven g-2 muon storage ring.}
\end{figure}

While it is clear that the current level of prediction creates a market
for improving on the CERN measurement, the predicted level of EWK
corrections at $151\times 10^{-11}$ is not all that much bigger than
the error of
$77\times 10^{-11}$ in the prediction, which is dominated by the hadronic 
uncertainty. Improving the low energy $e^+e^-$ measurements is a clear
concurrent goal.  The Brookhaven goal for measurement precision is
$\pm 35\times 10^{-11}$ for both $\mu^+$ and $\mu^-$.

\subsection{METHOD OF MEASUREMENT}

The charged pion decays by the weak charged current to $e\bar{\nu}_e$ or 
$\mu\bar{\nu}_{\mu}$.
The V-A form implies that the zero (or negligible)
mass of the neutrino forces helicity selection. To conserve
angular momentum in the decay of the zero spin pion, the massive 
charged lepton must be in the helicity state disfavored
by V-A. Since the muon is much heavier than the electron, 
the electron channel is much more suppressed by helicity,
$\sim10^{-4}$.  The muons
are longitudinally polarized.  So if you have a beam of pions, muons from
forward decay have the least momentum change and will most likely remain
in the beam. Thus, one can readily obtain beams of longitudinally polarized
muons. The basic method of the experiment is to put polarized muons in
a magnetic field and measure their spin precession.  The highest momentum 
electrons in the 
decay $\mu\rightarrow e\bar{\nu}_e\nu_{\mu}$ analyze the muon spin.

The experiment is done by putting muons into a storage ring. Lots of physics
results come from various storage rings, but the g-2 ring, shown in Fig.
1, is rather different.  For a precision measurement, you need as well known
and as uniform a magnetic field as possible. Thus, the ring is a continuous 
bending magnet.
The injection needs to be carefully done to avoid messing that up.
The time scale for storage is short due to the muon lifetime, so imbedded
electrostatic quadrupoles are sufficient to capture the beam.  Decay
electrons are detected in stations at the array of windows. 

\begin{figure}
\vspace{.1in}
\hspace{.2in}\epsfysize2.2in\epsffile{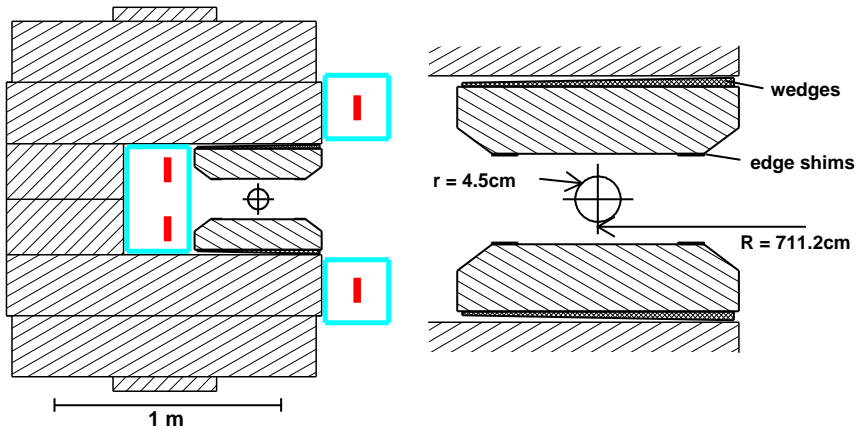}
\caption{The profile of the g-2 muon storage ring magnet. The boxes are
the vessels for superconducting cable.  The detail on the right shows
allowance for field adjustment.}
\end{figure}

\paragraph{Gedanken Problem.}
The g-2 ring uses an iron dominated superconducting 
magnet, which would not be
much stronger than typical conventional storage ring bending magnets.
The design momentum is 3.1 GeV/c. The SPEAR storage ring at SLAC, original
home of $\Psi$s, $\chi$s, $D$s, $\tau$s etc., has a design momentum of
4 GeV/c, yet it is something like five times larger than the g-2 ring. Why?
\vspace{.1in}

Injection is clearly an important problem. So far, they have used
pion decay. at the appropriate fortunate time, to give a slight momentum
kick for injection.  This is rather inefficient, and the pions which do
not decay run into things, creating a blast of stuff in the detectors, so that
measurements must wait till the detectors are stable.  A fast kicker is
in the works so that a muon beam can be directly injected. 

\begin{figure}
\vspace{.1in}
\hspace{.5in}\epsfysize4.5in\epsffile{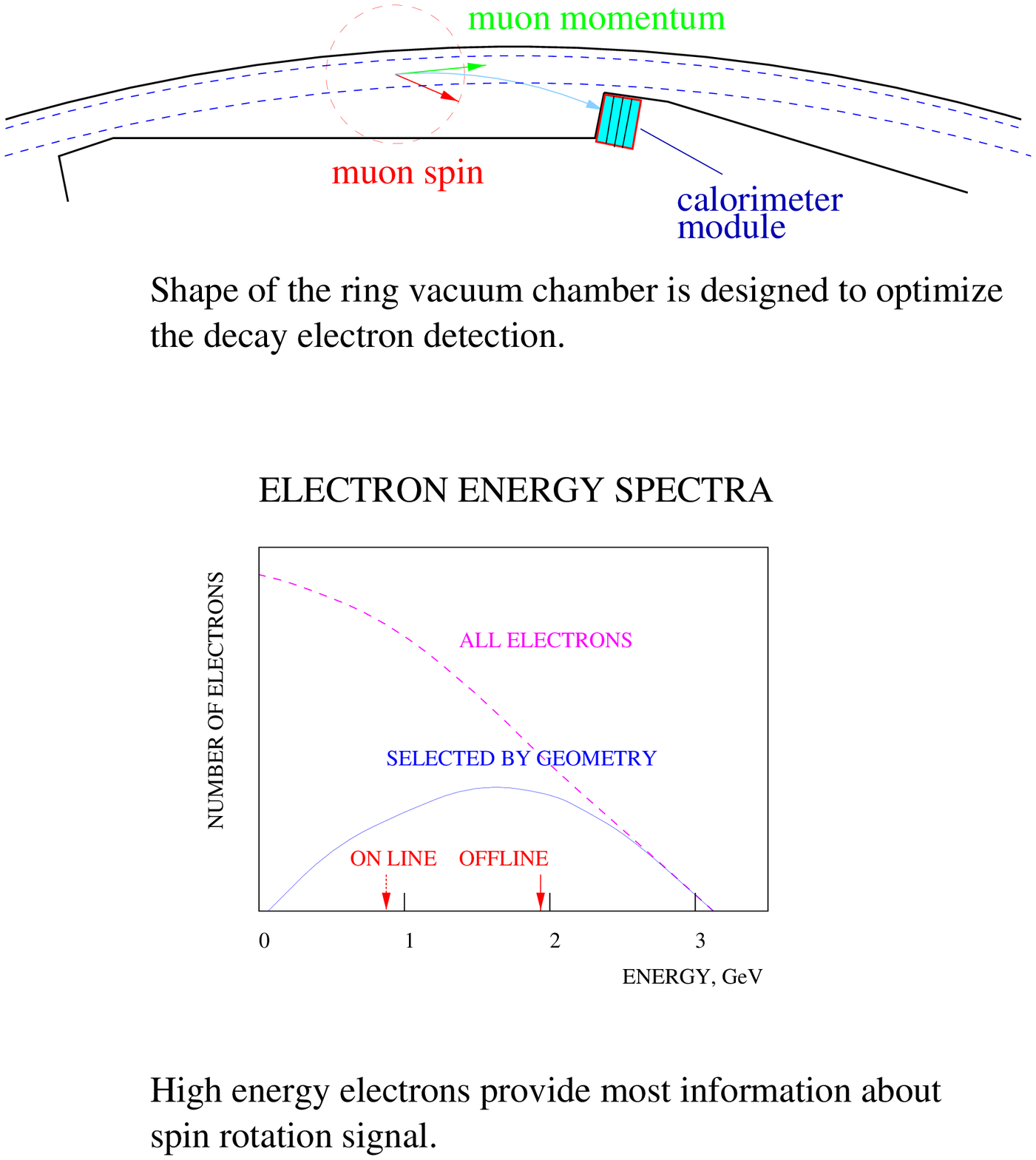}
\caption{Top: the decay electron window in the vacuum. Bottom: the energy
spectrum of detectable electrons with online and offline thresholds
marked.}
\end{figure}

Magnet quality is an ongoing project. The ring magnet profile is shown
in Fig. 2. Field mapping and orbit studies feed back to shimming the
iron pole tips. The profile and absolute value are both important; an
NMR probe provides the absolute field value by reference to the proton magnetic
moment.

\begin{figure}
\vspace{.1in}
\hspace{.5in}\epsfysize4.5in\epsffile[15 155 570 670]{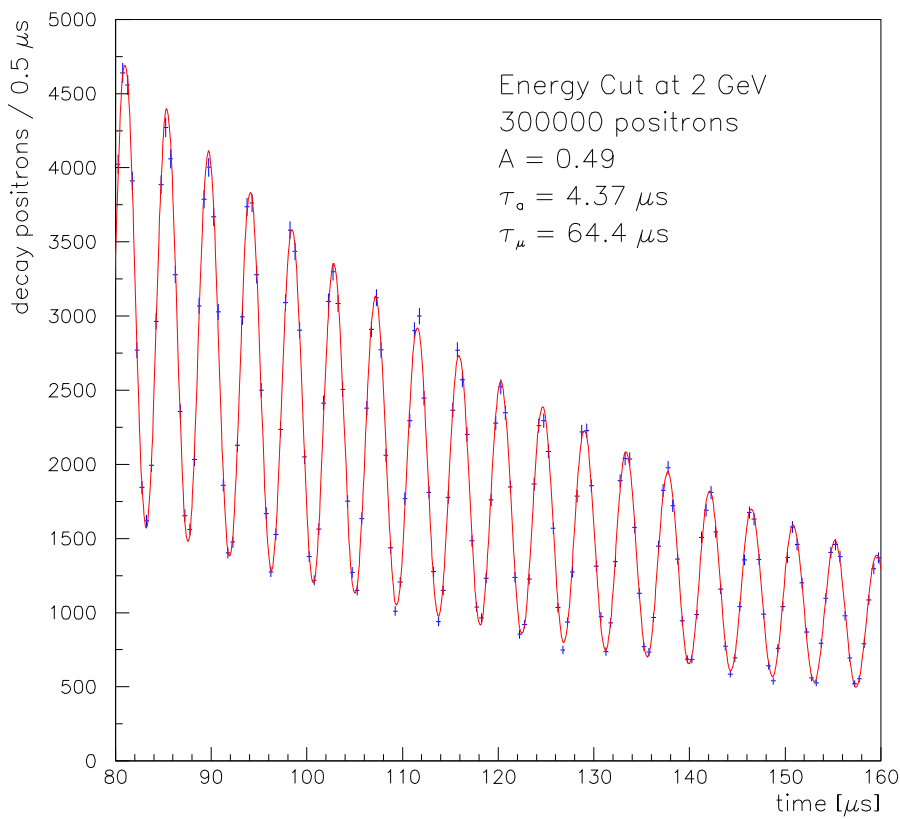}
\caption{The time spectrum of detected electrons. The period gives $a_\mu$.
They may eventually improve $G_F$.}
\end{figure}

\vspace{-.1in}
\paragraph{Gedanken Problem.}
The iron dominated superconducting magnet provides
good opportunity to control field quality. For the beyond LHC generation
hadron collider, VLHC or Eloiseatron, what are the tradeoffs between
high field (coil dominated) and iron dominated bending magnets?
\vspace{.1in}

What you want to measure is the difference in the cyclotron frequency
and the spin precession frequency given by\\
\vspace{-.1in}
\begin{center}
$\omega_a = {e \over mc}[a_{\mu}B - (a_{\mu} - 1/(\gamma^2 - 1))(B \times E)]$.\\
\end{center}
\vspace{.1in}
At Brookhaven, as at CERN, they choose the ``magic gamma'' for momentum
of 3.09 GeV/c, so the measurement is not sensitive to the electrostatic
quadrupole field.

The experimental measurement is illustrated in Fig. 3. The decay
electron exits a window on the inside of the ring, and is detected
with timing and energy in scintillating fiber calorimeters. The highest
energy electrons are most correlated to spin direction. Thus, understanding
the selection threshold imposes constraints on calibration and the 
stability of the calorimeter.

\vspace{-.1in}
\paragraph{Gedanken Problem.}
Consider the general cases, as described in J. Virdee's lectures, of
calibration systems for calorimeters
contrasting setup for data taking and {\it in situ} maintenance.

\subsection{RESULTS AND PROSPECTS}

The measurement is illustrated in Fig. 4, measuring\cite{Hughes} \\
\vspace{-.1in}
\begin{center}
$a_{\mu^+} = 116592500 (1500) \times 10^{-11}$.\\
\end{center}
\vspace{.1in}
The level of the statistical error is 
approaching CERN. The systematics are at the level of 
$\sim\pm 300\times 10^{-11}$, including magnetic field systematics and
detector related uncertainty.

Work is going into various improvements. Muon injection will give a 
big statistical boost. 
Field improvements with shims and trim coils will improve systematics, as will
improved detector understanding and
tracking of the electrons to verify where the muon beam is.  And, of course,
the hadronic correction prediction is also getting attention.

In 1998, they expect to get to the level of $\pm 115\times 10^{-11}$ for
$\mu^+$. Full design precision for both muon charges should be realized in
2002.

\section{NuTeV}

\subsection{GOALS OF THE MEASUREMENT}

NuTeV is the latest incarnation of the CCFR (Chicago, Columbia, Fermilab,
Rochester) neutrino experiment at Fermilab. They were off looking for
wrong-sign heavy leptons when Gargamelle observed neutral currents.
Their confirmation of neutral currents 
at the 1974 London Rochester conference silenced the
many vocal skeptics, thus contributing to the recognition of
electroweak unification.

The basic measurement then and now is to measure the relative rate of
neutral current (NC) and charged current (CC) interactions, measuring the
on-shell weak mixing angle\\
\vspace{-.1in}
\begin{center}
$sin^2\Theta^{on-shell}_W \equiv 1 - {m(W)^2 \over m(Z)^2}$.\\
\end{center}
\vspace{.1in}
The experiment uses muon neutrinos and distinguishes
NC from CC by the absence of penetrating muons in the final state.
Over several generations of such experiments, systematics such as
detailed understanding of the boundaries of the detectors, and more importantly,
the charm quark mass, have limited the measurement, as seen in Fig. 5.
Neutrinos can interact with a strange sea quark, producing charm; the
charm decay can give a muon, complicating the measurement. The charm mass
is needed to predict the CC rate.  Two muon events can be used to get some
measure, but this seems to have plateaued at the level, in the appropriate
QCD order, of $m_c \sim1.3\pm0.3$ GeV/c$^2$. Another important systematic
effect is the level and understanding of $\nu_e$ in the beam, as electron
CC  look like muon NC. Electron neutrinos from decays of
$K^0_L$ from the production target are a problem due to significant
uncertainties in the $K^0_L$ production rate.

\begin{figure}
\vspace{.1in}
\hspace{.2in}\epsfysize3.2in\epsffile{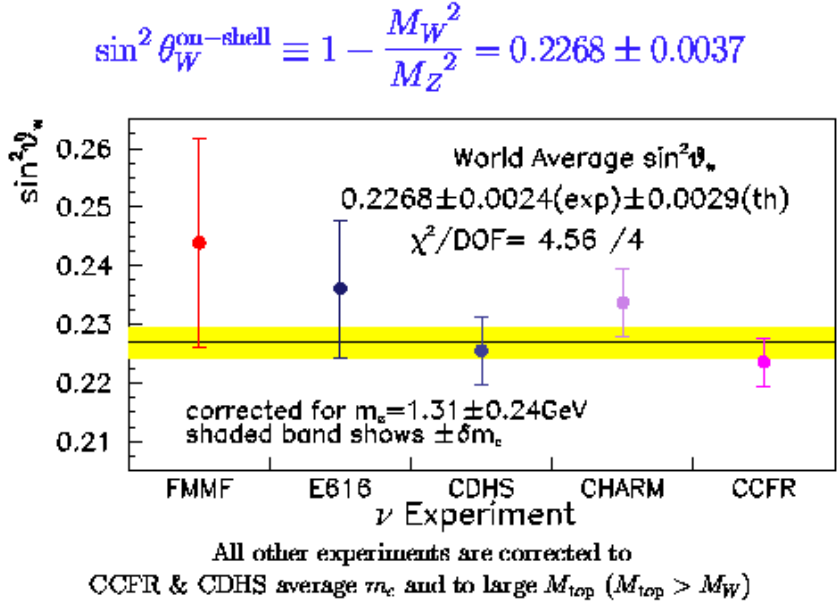}
\caption{The evolution of neutrino weak mixing measurements showing
the charm systematic limit.}
\end{figure}

The basic goal of the NuTeV measurement is to escape these systematic
limits.  By creating a beam with very little neutrino/antineutrino
cross talk, it can use the Paschos-Wolfenstein relation\cite{pw}\\
\vspace{-.15in}
\begin{center}
($\sigma_{NC}-\bar{\sigma_{NC}})/(\sigma_{CC}-\bar{\sigma_{CC}})
= \rho^2(1/2 - sin^2\Theta_W)$,\\
\end{center}
\vspace{.05in}
obtaining weak mixing from the difference of neutrino and antineutrino
cross sections.
Thus, the sea cancels and the remaining difference in charm production, 
due to valence $d$ quark, is Cabbibo suppressed. The goal is
to make a competitive inference of the $W$ mass.

\subsection{METHOD OF MEASUREMENT}

The CCFR detector consists of 690 tons of tracking calorimeter with
transverse square planes of iron, liquid scintillator,
and drift chambers. The central 390 tons are considered fiducial.
This is followed by an extensive system of drift chambers and iron toroids
to measure muon momentum. The basic measurements are illustrated with
event pictures, see Fig. 6.

\begin{figure}
\vspace{.1in}
\hspace{.2in}\epsfysize2.1in\epsffile{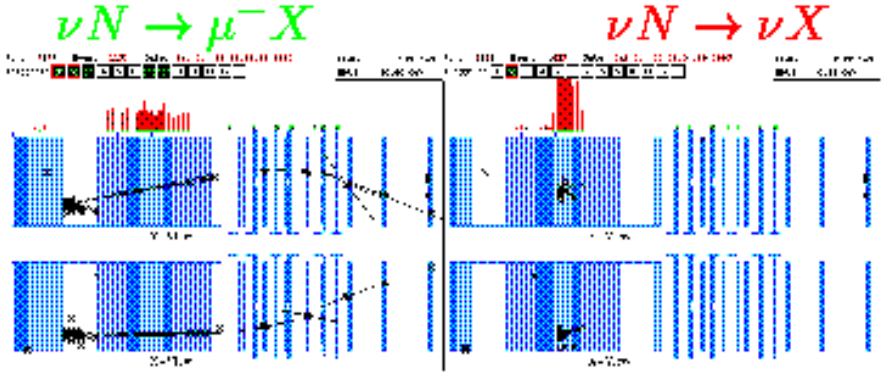}
\caption{Neutrino interactions in NuTeV/CCFR, left CC and right NC.}
\end{figure}

The most significant change for NuTeV, compared to CCFR, is the beam.  Neutrino
experiments typically use zero degree production to optimize yield, using
magnetic horn focusing 
for broad-band beams, or beam line (quadrupole) focusing for
narrow-band beams. The NuTeV beam, shown in Fig. 7, 
looks away from zero degrees, avoiding sensitivity
to upstream scraping, reducing the contribution from $K^0_L$ decays, 
and thoroughly removing wrong sign particles. 
The usual decay region is followed by muon shielding, then the detector.
Thus the beam should be
well understood, as is illustrated in Fig. 8.  Note the two peaks
corresponding, left-to-right,  
to neutrinos from pion and kaon decays. Residual beam systematics are dominated
by the $K^{\pm}$ e3 branching ratio uncertainty. 
The flux can be studied with quasielastic CC events,
$\nu p \rightarrow \mu n$ with perhaps nuclear breakup, but not pion
production.

\begin{figure}
\vspace{.1in}
\hspace{.2in}\epsfysize2.5in\epsffile{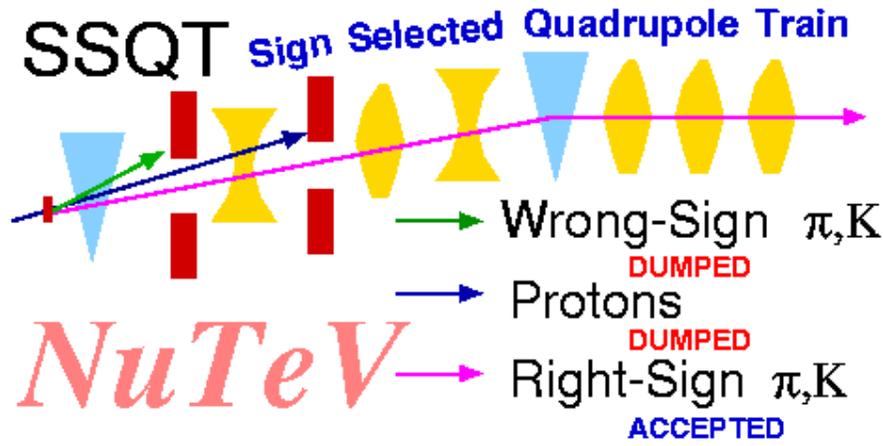}
\caption{NuTeV production target and beam line into the decay region.}
\end{figure}

\begin{figure}
\vspace{.1in}
\hspace{.6in}\epsfysize4.0in\epsffile{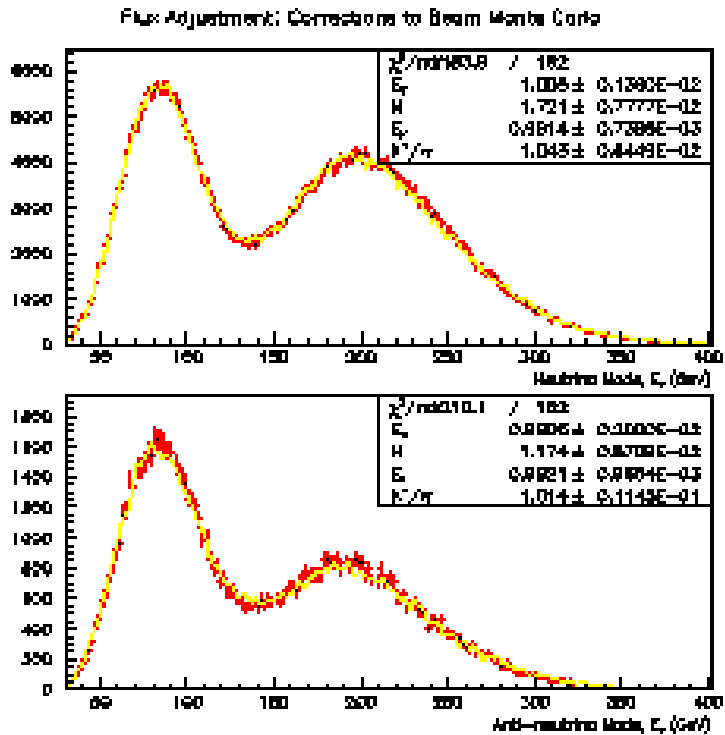}
\caption{NuTeV neutrino and antineutrino flux, detected and MC.}
\end{figure}

\vspace{-.1in}
\paragraph{Gedanken Problem.}
Consider the general problem of high intensity production targeting -
how would you optimize for producing antiprotons, muons, or tau neutrinos?
\vspace{.1in}

The basic measurement consists of measuring the length of events. Length is 
measured in units of
counter planes, which are 
separated by 10 cm of iron. The boundary between short
and long is taken at 20 planes.  A concurrent test beam line allows the
detector, including boundaries where muons could sneak out, to be
well understood. Convolving this understanding with the flux 
in Monte Carlo simulation should
allow a detailed understanding of length distributions as illustrated
in Fig. 9. The region of the cut is reasonably well described, as shown in
the insets, and the ratio
short/long can be unfolded to give NC/CC. 

\begin{figure}
\vspace{.1in}
\hspace{.5in}\epsfysize4.0in\epsffile{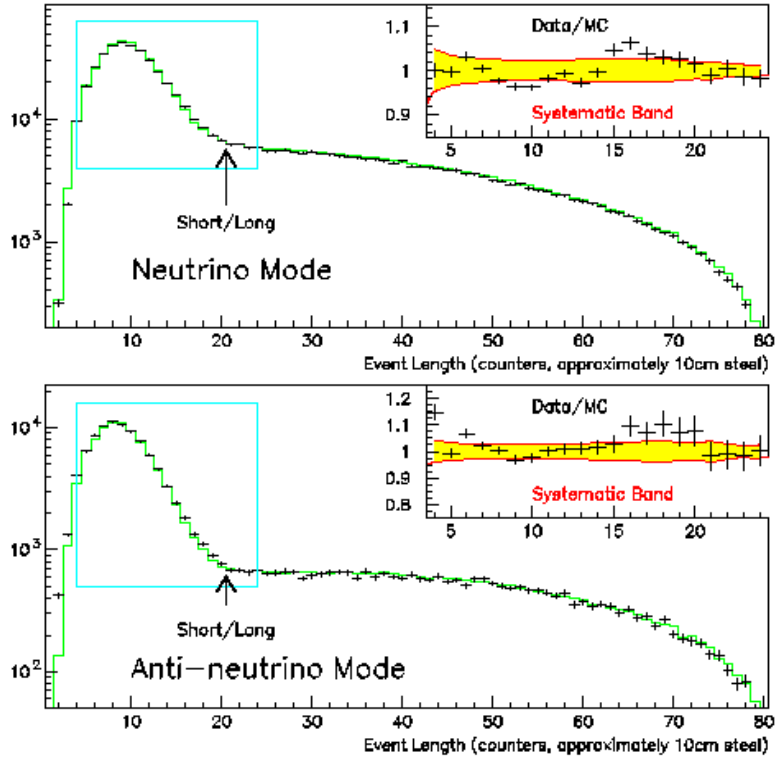}
\caption{NuTeV neutrino and antineutrino events as a function of
event length. Insets are data/MC ratios in the region of the cut.}
\end{figure}

\subsection{RESULTS AND PROSPECTS}

The NuTeV run was completed in 1997, and preliminary results have been
given.\cite{ksmf} The statistical precision for
$sin^2\Theta_W$ is $\pm0.00190$, comparable to the 
CCFR result.  
The physics model systematics level is $\pm 0.00070$, with detector
systematics at $\pm 0.00075$. Unlike CCFR, the NuTeV measurement is
statistically dominated.  Using the measured top mass, the NC/CC ratio 
gives $sin^2\Theta_W = 0.2253 \pm 0.0022$, which corresponds
to $m(W) = 80.26 \pm 0.11$ GeV/c$^2$. Combining NuTeV with the CCFR result
gives\\
\vspace{-.1in}
\begin{center}
$sin^2\Theta^{on-shell}_W = 0.2255 \pm 0.0021$.\\
\end{center}
Their goals have been largely achieved.
Some improvement in experimental systematics may come with further
analysis. No further data taking is planned.

\section{Digression on Collider Detectors}

\begin{figure}
\vspace{.1in}
\hspace{.2in}\epsfysize4.0in\epsffile{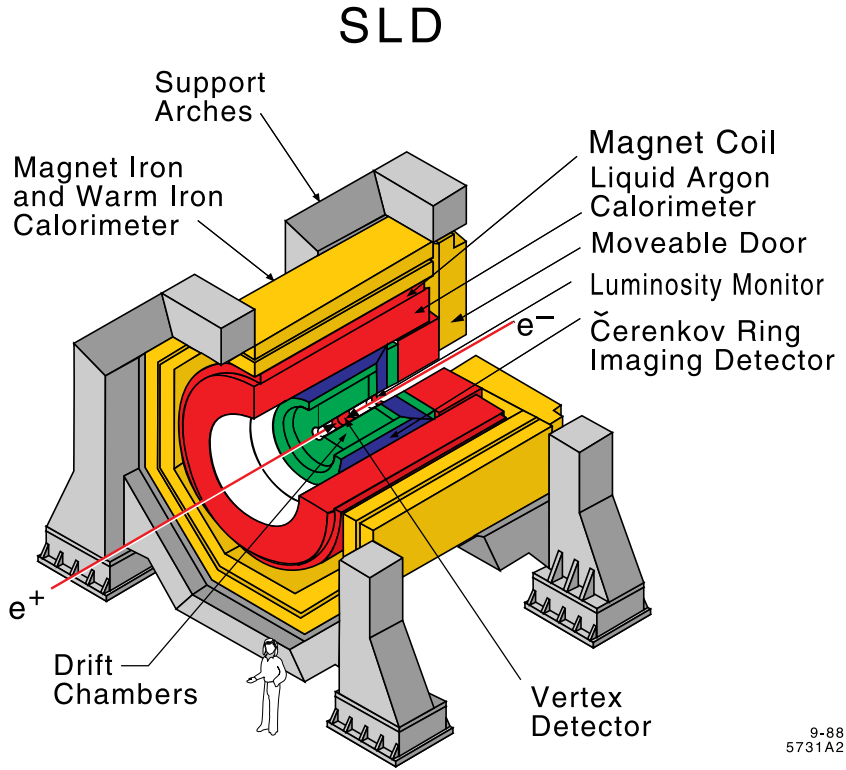}
\caption{SLD as an example of a generic collider detector. The warm iron
calorimeter doubles as a muon detector.}
\end{figure}

The experiments remaining are at either $e^+e^-$ or $\bar{p}p$ colliders.
Almost all such collider detectors, as well as those at HERA, are quite
similar.  Good examples, ATLAS and CMS for LHC, were thoroughly described
in the
lectures of F. Pauss.  A tracking volume around the interaction is defined
inside a solenoid. These days drift chambers for magnetic tracking are
complemented, as discussed
in the lectures of I. Abt, by silicon detectors. There may also be particle
ID. The tracking volume is 
surrounded by calorimeters, and a muon identification
system surrounds that. This is illustrated by SLD, shown in Fig. 10.

Electrons are identified by matching a track to suitable calorimeter 
measurements.
Muons match a track inside to a track or track stub outside.  To identify
$\tau$ leptons, one demands
one or three tracks corresponding 
to a relatively narrow calorimeter energy cluster, isolated from other 
activity. In the case of hadron
colliders, the detectors must make ID information available for fast
triggers.  

The silicon detectors are used to identify $b$ quarks by
observing secondary vertices.  This procedure is usually marginal for charm,
and $c$ quarks tend to be identified by reconstructing exclusive $D$
decay modes or using the soft pion from $D^*\rightarrow\pi^{\pm} D$ with 
partial
reconstruction of the $D$ to show the low Q for the pion. Clearly, $b$ and $c$
identification are correlated; charm decay is only three times faster, and 
bottom decays produces charm.

\vspace{-.1in}
\paragraph{Gedanken Problem.}
How would you trigger, at a hadron collider, on $\tau$ leptons? Has this
worked?
\vspace{.1in}

The first detector of this form was Mark I at SPEAR in the early 70s.
CDF at the Tevatron was the first such for hadron colliders. Notable
exceptions with no central field 
are the Crystal Ball, UA2, and pre-upgrade D\O. 
UA1 had a dipole magnet.

\section{LEP1 $Z$ Studies}

\subsection{GOALS OF THE MEASUREMENT}

The basic goal of the LEP1 program was a comprehensive study of the
$Z$ boson.  The $Z$ is observed as an s channel resonance in $e^+e^-$
collisions.  The important issues here are the $Z$ lineshape,
basic aspects of decays, and in particular the charge and $\tau$ polarization
asymmetries which result
from parity violation.
The presence of both vector 
and axial vector coupling gives asymmetries which measure
weak mixing. 
A definition of the weak mixing angle
$sin^2\Theta_{W eff}$ is used which avoids loss of
precision to uncertain top and Higgs mass dependence.
For leptons, this is simply\\
\vspace{-.1in}
\begin{center}
$sin^2\Theta^{lept}_{eff} \equiv 1/4(1 - g_{V\ell}/g_{A\ell})$. \\
\end{center}
\vspace{.1in}
Checking the electroweak radiative corrections was one of the central
original motivations for the LEP program.

\subsection{METHOD OF MEASUREMENT}

Four detectors, Aleph, Delphi, L3, and Opal, use the LEP collider, shown
in Fig. 11.  The detectors need to identify the interactions with definable
efficiencies, and tell the $\pm e^+$ direction. One concern is to measure the
absolute luminosity; small high precision calorimeters are used to count
small angle Bhabha ($e^+e^-$ elastic) scattering. The experimental method
has evolved sufficiently that the luminosity uncertainty is dominated by
the QED calculation.\cite{leplum}

\begin{figure}
\vspace{.1in}
\hspace{.7in}\epsfysize4.0in\epsffile[65 130 535 690]{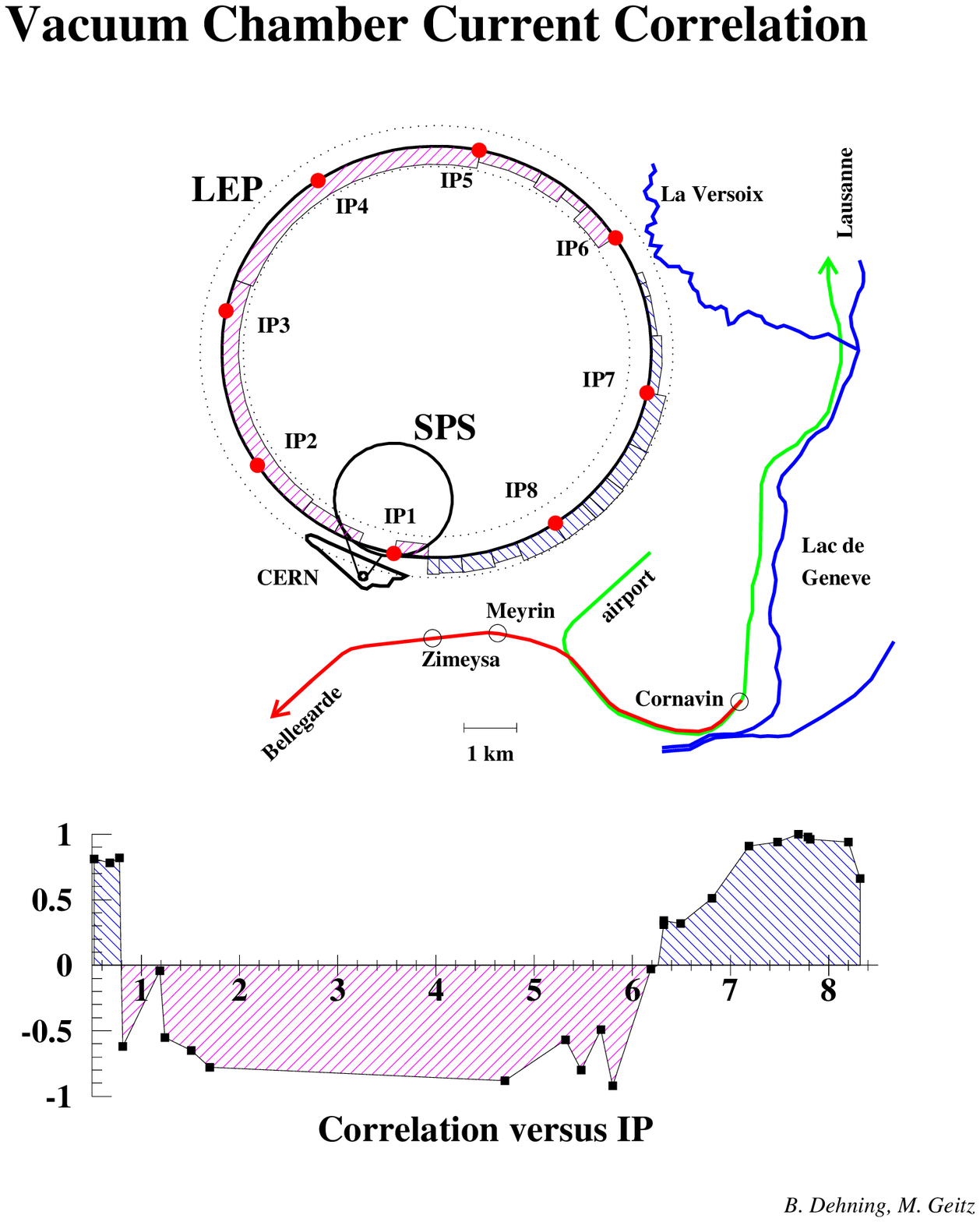}
\caption{The LEP ring showing proximity to Lake Geneva and electric
rail lines. Also shown is ground current in the beam pipe. L3, Aleph, Opal
and Delphi are at IPs 2, 4, 6 and 8. The circumference is 27 km.}
\end{figure}

A greater concern is the absolute beam energy calibration, which determines
the precision of the $Z$ mass measurement.  Survey and field mapping, as for
g-2, is not good enough by itself. The basic measurement uses
resonant depolarization.\cite{lepecal} With favorable accelerator parameters
avoiding resonances, synchrotron radiation will tend to polarize the beams. 
Applying transverse RF, with precisely the correct frequency, causes the 
polarization to be lost.  The intrinsic width for this procedure corresponds
to 200 KeV. One corrects for RF energy gain and synchrotron radiation loss
as appropriate from the measurement to each interaction point.

\begin{figure}
\vspace{.1in}
\hspace{.7in}\epsfysize4.0in\epsffile[50 85 500 690]{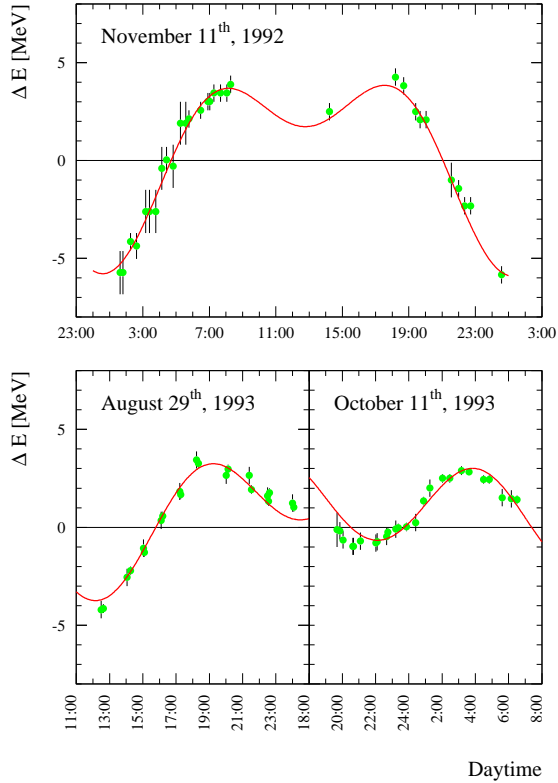}
\caption{The LEP beam energy correlation to the moon, revealing the effect
of tides.}
\end{figure}

In practice, during energy scans, this measurement is made at the end of
stores.  The measurements show a 5 MeV spread in beam energy predicted 
from field and orbit measurements minus resonant depolarization values.
If the depolarization
measurement times are representative, one can combine them statistically
to get absolute energy.
But precision and confidence are improved if one can understand and model
the time dependence which causes the spread.

The LEP program is clearly big science.  The first correlation found
was with the phase of the moon, shown in Fig. 12. The tidal expansion and
contraction of the LEP ring is noticeable.  Another influence on the size
of the LEP ring is the amount of water in the ground; this is fairly 
well modelled by the water level of Lake Geneva, as see in Fig. 13.
These correlations gave reasonable confidence in measurements through 1994.

\begin{figure}
\vspace{.1in}
\hspace{.2in}\epsfysize4.0in\epsffile{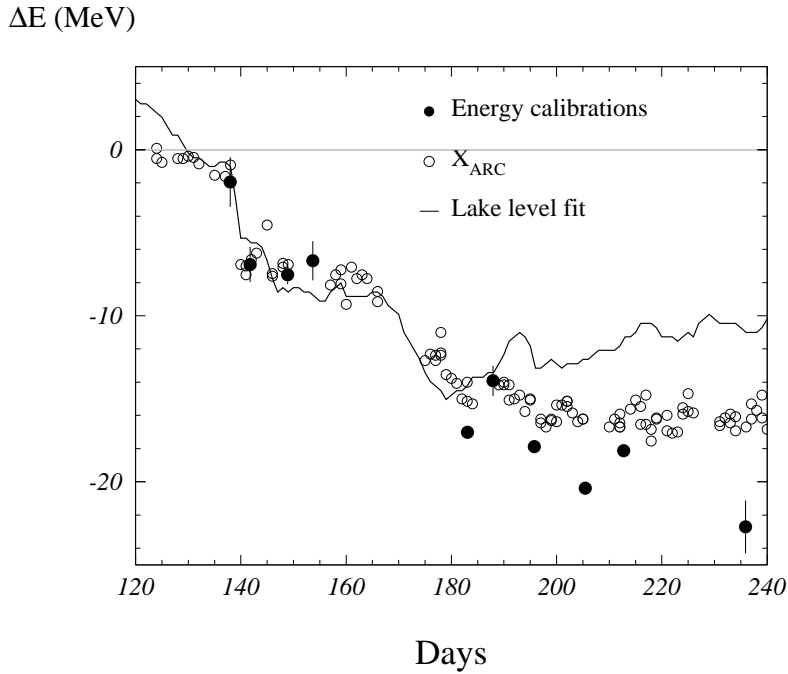}
\caption{The LEP beam energy correlation to the water level of Lake Geneva.}
\end{figure}

A major energy scan was undertaken in 1995. During the run, unexpected
time dependences were found in NMR readings. The patterns were found to be
reasonably regular with daytime. Eventually these were found to be
correlated with running of the TGV trains, as seen in Fig. 14.  Apparently
the TGV actually uses the ground to return current, and significant
levels of current were found in the LEP beam pipe, shown in Fig. 11.
A time dependent correction was determined, and a retroactive correction
applied to previous data.

\begin{figure}
\vspace{.1in}
\hspace{.7in}\epsfysize4.0in\epsffile{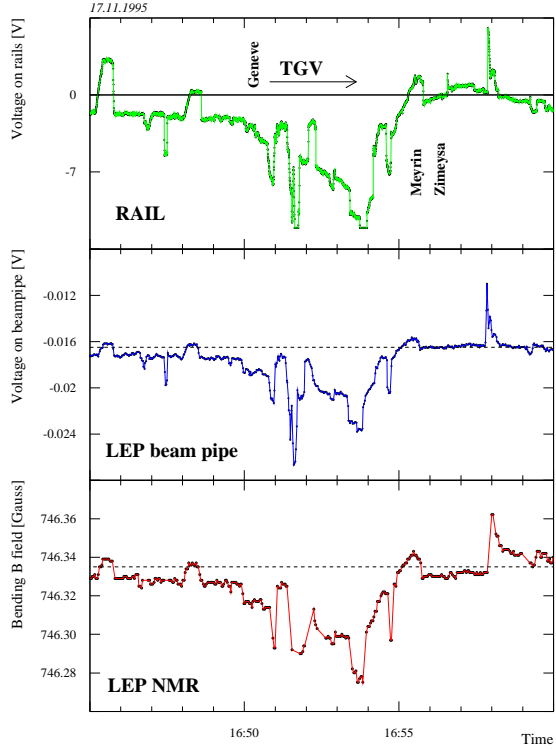}
\caption{The LEP beam energy (NMR) correlation to current in the
beampipe and power to the TGV.}
\end{figure}

Charge asymmetries are straightforward to measure, although physics
interpretation for hadrons can involve considerable sophistication in QCD and
in relating
jets to quarks and understanding charge correlations and efficiencies.
The polarization of $\tau$ leptons is analyzed in the decay modes
$\pi\nu$ and $\rho\nu$, as well as $a_1\nu$ and leptonic decays.
The polarization angular distribution varies as\\
\vspace{-.1in}
\begin{center}
$P_{\tau}(cos\theta) = (A_{\tau}(1+cos^2\theta)+2A_ecos\theta)
/(1+cos^2\theta+2A_{\tau}A_ecos\theta)$,\\
\end{center}
\vspace{.1in}
where $A_f \equiv 2g_Vg_A/(g_V^2+g_A^2)$ for $Z f$ coupling.

\subsection{RESULTS AND PROSPECTS}

The L3 version of the $Z$ scan is shown in Fig. 15. I will quote combined
results from Vancouver 98.\cite{LEPEWWG} The lineshape is well described by\\
\vspace{-.1in}
\begin{center}
$m_Z = 91.1867 (21)$ GeV/c$^2$\\
\vspace{0.05in}
$\Gamma_Z = 2.4939 (24)$ GeV.\\
\end{center}
\vspace{.1in}
With efficiencies and luminosity normalization, 
one obtains a peak cross section
of 41.491 (58) nb and an average leptonic decay width of 83.91 (10) MeV.
So there are three flavors of neutrinos, and no significant rate to any channel
we don't know about.

\begin{figure}
\vspace{.1in}
\hspace{.9in}\epsfysize4.0in\epsffile[38 100 490 715]{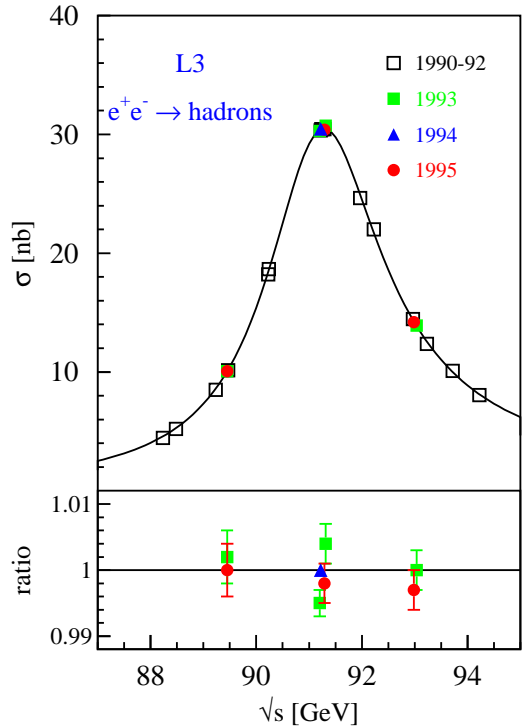}
\caption{The LEP $Z$ lineshape as seen by L3.}
\end{figure}

The $Z$ asymmetry measurements are summarized in Table 2. 
The leptonic
measurements are a combination of $\mu^+\mu^-$ (most precise), with
$e^+e^-$ (diluted by t channel exchange), and $\tau^+\tau^-$ 
(lower efficiency). Of hadronic final states, the best measurement
is for $b$ quarks,
as the tag tends to tell you where the quark really is. Identified 
charm and jet charge, which
represents light quarks, have similar precision.  
Clearly, there are correlations among the hadronic measurements.

\begin{table}[htb]
\begin{center}
\caption{LEP measurements of $sin^2\Theta_{W eff}$ from $Z$ asymmetries.}
\begin{tabular}{lrr}
\hline
Final state & Value & Error \\
Leptons & 0.23117 & 0.00054 \\
Jet charge & 0.23210 & 0.00100 \\
$b$ jet & 0.23223 & 0.00038 \\
$c$ jet & 0.23200 & 0.00100 \\
$\tau$ pol $A_e$ & 0.23141 & 0.00065 \\
$\tau$ pol $A_{\tau}$ & 0.23202 & 0.00057 \\
\hline
Average & 0.23187 & 0.00024 \\
\hline
\end{tabular}
\end{center}
\end{table}

\vspace{-.1in}
\paragraph{Gedanken Problem.}
Using several samples to make a given measurement, where there is
cross talk between the samples, is a general problem.  How do you deal
with it?
\vspace{.1in}

The $\tau$ pair decay polarization 
angular distribution 
terms, which depend on electron and on tau coupling, are listed
separately. All the asymmetry measurements are reasonably consistent;
split out as listed, the $\chi^2$ per degree of freedom is 3.2/5.

The LEP1 data will not be significantly increased; the statistical impact
of the occasional LEP2 $Z$ calibration run is insignificant.
Most analyses are
reasonably mature.  Some updates and refinements may be expected.

\section{SLC $Z$ Studies}

\subsection{GOALS OF THE MEASUREMENT}

The goal of SLC is to use initial state electron polarization to give the
most precise possible measurement of the $Z$ asymmetry, and thus weak
mixing. An overriding goal, not relevant here, is to demonstrate the
feasibility of linear colliders.

\subsection{METHOD OF MEASUREMENT}

To measure the $Z$ asymmetry $A_{LR}$, one simply measures the cross
section difference for left- and right-handed electrons at the $Z$ pole.
All the detector needs to do is count events $(N_L - N_R)/(N_L + N_R)$,
then divide out the beam polarization. The result depends
on $sin^2\Theta_{W eff}$ as\\
\vspace{-.1in}
\begin{center}
$A_{LR} = 2(1 - 4sin^2\Theta_W)/(1 + (1 - 4sin^2\Theta_W)^2)$.\\
\end{center}
\vspace{.1in}
The tricky part is obtaining and understanding the beam polarization.

\begin{figure}
\vspace{.1in}
\hspace{1.5in}\epsfysize3.75in\epsffile{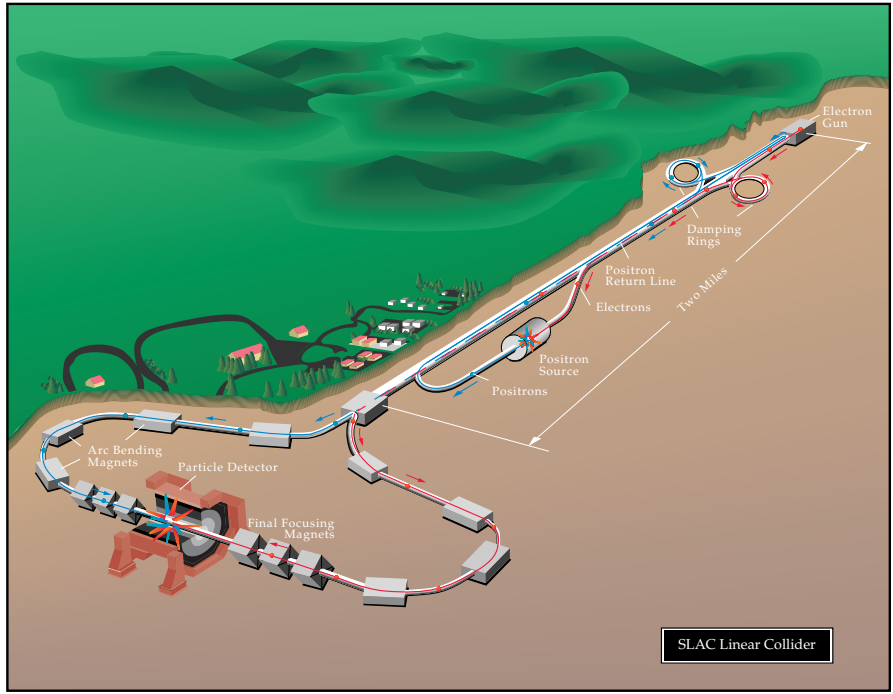}
\caption{The linear collider at SLAC. The linac is 3.2 km long.}
\end{figure}

SLC is shown in Fig. 16.\cite{rowson} 
Positrons are made
by accelerating electrons down most of the length of the linac, 
hitting a target from which positive charges are collected and returned
to the start of the linac. After some acceleration, they are put into
a damping ring.

The polarized electron source was a great technical triumph. It produces
right- or left-handed electrons based on a random number. These are accelerated
a bit, then transferred, with spin rotated to vertical, to a damping ring. 
The damping rings are synchrotron radiation cooling rings, compressing
the phase space of the electrons and positrons to allow a small beamspot
and high luminosity. 

For a pulse of the collider, the bunch of 
electrons, spin rotated back to longitudinal,
are followed by the positron bunch, accelerated down the linac.
At the end of the linac the electron 
polarization is rotated back to vertical for the 
arc, and checked with a Moller polarimeter. Preserving the polarization
through the arc is also tricky, requiring delicate alignments. The polarization
is rotated back to longitudinal going into the final focus and IP. The electron
polarization is measured downstream of the IP with a Compton polarimeter,
shown in Fig. 17; a Cerenkov detector array at several angles measures
the asymmetry for recoil electrons which scatter off circularly polarized laser
light.

\begin{figure}
\vspace{.1in}
\hspace{.4in}\epsfysize3.6in\epsffile{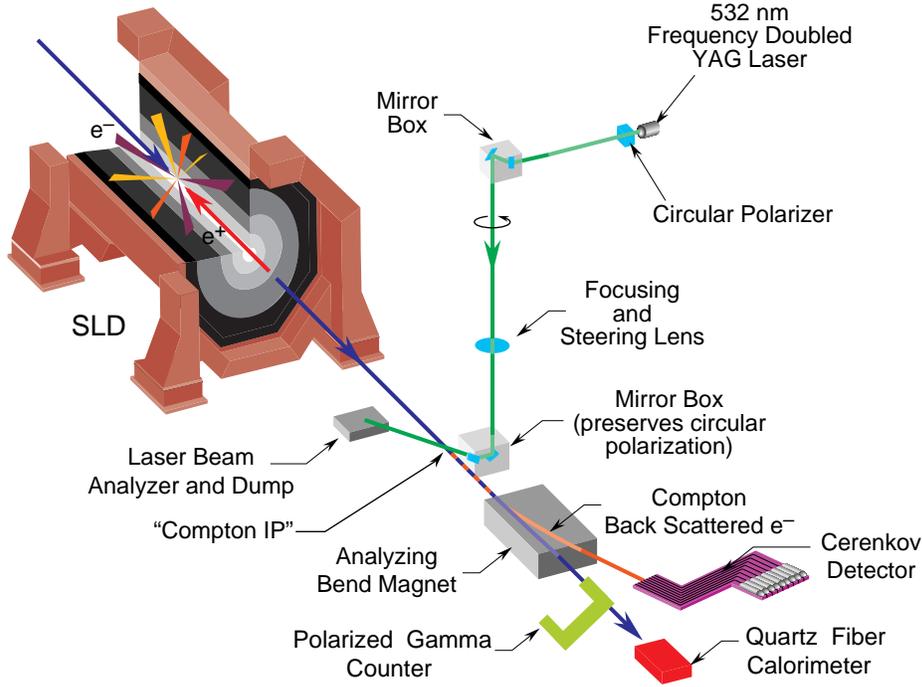}
\caption{The Compton polarimeter.}
\end{figure}

A litany of systematic effects has been investigated, including laser
polarization, noise, beam optics at the IP, bunch tails, and possible
positron polarization. Everything seems to be 
under control. Average polarization
is $77.25 \pm 0.52\%$.

\subsection{RESULTS AND PROSPECTS}

The result, as of Vancouver 1998,\cite{baird} is 
$A_{LR} = 0.1510 \pm 0.0025$;
where the error includes $\sim\pm 0.0010$ systematics in quadrature.
This gives $sin^2\Theta_{W eff} = 0.23101 (31)$. This is indeed the
most precise single measurement, 
and it continues the historic trend of being noticeably
lower than other determinations. 
In combination with the LEP measurements one gets\\
\vspace{-.1in}
\begin{center}
$sin^2\Theta_{W eff} = 0.23155 (19)$.\\
\end{center}
\vspace{.1in}
With LEP itemized as listed in Table 2,
this corresponds to $\chi^2$ per degree of freedom of 8.1/6.
So, in principle, a PDG $S^*$ factor for combining measurements should
be used to increase the combined error by $\sim\times 1.15$. In previous
years the consistancy has been considerably worse, and in practice even the
PDG struggles dealing with it.\cite{PDG} 
 
SLC/SLD has just completed running. They have some data not
included above, and some systematic studies continue. A final update may
come during 1999.

\section{The Tevatron $W$ Mass}

\subsection{GOALS OF THE MEASUREMENT}

For a while, the only way to study $W$ and $Z$ bosons was at the
$S\bar{p}pS$ collider at CERN. Parts of that program have continued at
the Tevatron collider at Fermilab, remaining competitive in the LEP2 era.
These include rare $W$ decay searches, the $W$ width, and of note here,
the $W$ mass. Each of the Tevatron collider experiments, CDF and D\O,
would like to measure the $W$ mass, with existing data, 
as well as any one LEP experiment
will, $\sim\pm 100$ MeV/c$^2$. Eventually after 
running with the ongoing upgrades
complete, each hopes to improve on the final LEP2 precision.

\subsection{METHOD OF MEASUREMENT}

When detecting $e^+e^-$ collisions at the $Z$ pole, if you see particles
which vaguely correspond in energy to twice the beam energy, that is a $Z$.
Vector bosons are readily produced in hadron collisions, but come 
in association with
soft particles, which I will call ``X'' as in $\bar{p}p \rightarrow W + X$. 
Generally, $Z$ and $W$ bosons are observed at hadron colliders 
in leptonic decays. 

X is made of low transverse momentum (few hundred MeV/c$^2$) 
particles which evenly occupy longitudinal
phase space, pseudorapidity $\eta = -ln(tan(\theta/2))$. For high
luminosity, the inelastic cross section is large enough that extra
``minimum bias'' inelastic events overlap the events of
interest.  These events consist of soft particles distributed evenly
in $\eta$. For the
Tevatron running six bunches, 3.5 microseconds between bunches, 
an average of one overlap event corresponds
to a luminosity of $\sim6 \times 10^{30} cm^{-2} sec^{-1}$. The most recent
Tevatron data had an average luminosity of almost twice that. As luminosity
goes up, so does X. For the high luminosity upgrade, the luminosity will be
spread out in more bunches, which
will keep X down. At LHC, the bunches are 25 ns apart, but X
gets rather large for luminosity anywhere near design.

$W$ and $Z$ production is described, 
as discussed by J. Stirling, by PDFs giving
probabilities for finding partons in the proton and antiproton, and the 
Drell-Yan quark annihilation process.\cite{dy} 
QCD resummation\cite{resum} can be used to define parameters to
describe $p_T(W)$ in the low $p_T(W)$ region relevant to measuring the $W$ 
mass. These parameters are determined or checked
by measuring $p_T(Z)$. If $p_T(W)$ were perfectly predicted,
then the $p_T$ of the decay charged lepton would be used directly to estimate
the $W$ mass, minimizing the effect of X.

As pointed out in Stirling's discussion of PDFs, the momentum fraction
distributions for
${u}$ and ${d}$ in the proton are different in shape. As $W$s
are produced $u\bar{d} \rightarrow W^+$ and charge conjugate, there is a net
$W$ charge correlation, with $W^+$ tending to be produced with net 
longitudinal momentum along the proton direction. For central values of
$\eta$, this charge 
correlation applies to the decay lepton, as seen in Fig. 18.\cite{wasym}
For higher absolute $\eta$, parity violation in the decay reverses the 
correlation seen in the decay leptons. 
Measuring this charge asymmetry gives a  PDF
constraint that is quite useful for measuring the $W$ mass.

\begin{figure}
\vspace{.1in}
\hspace{.2in}\epsfysize3.6in\epsffile[15 195 525 630]{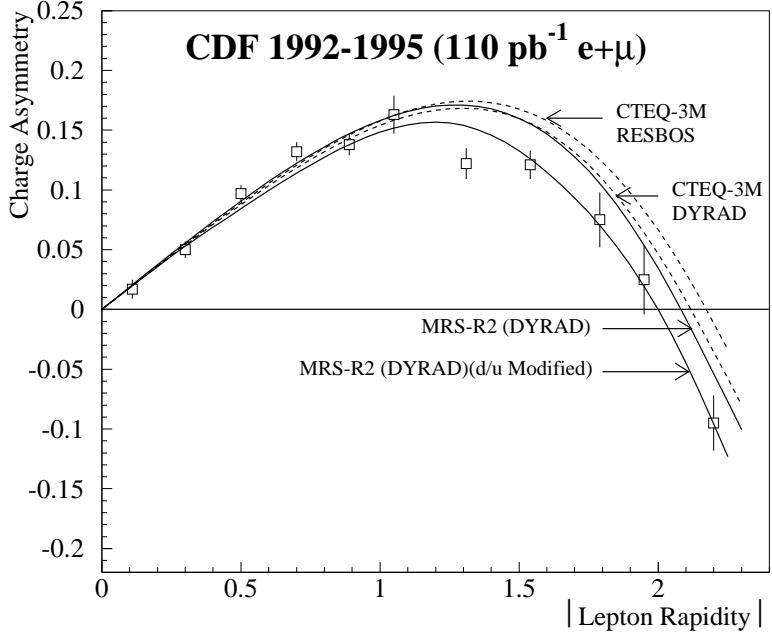}
\caption{The CDF measurement of $W$ lepton charge asymmetry. CTEQ and MRS
are collaborations which do general fits for PDFs. DYRAD and
RESBOS are NLO and resummed Monte Carlo generators respectively.}
\end{figure}

While the net longitudinal energy flow is essentially unmeasured, due to
the finite longitudinal acceptance, the net transverse energy is reasonably
well measured for modest X. The net calorimeter energy flow, other than
the lepton, is the sum of response to the hadronic recoil against $p_T(W)$
and X. The neutrino is inferred from the net calorimeter imbalance. 
Typical $W$ event selection requires lepton and missing $E_T$ above 20
GeV.

The optimal strategy for the $W$ mass uses transverse mass,\\
\vspace{-.1in}
\begin{center}
$m_T \equiv 
\sqrt{(E_T(\ell) + E_T(\nu))^2 - (\vec{p}_T(\ell)+\vec{p}_T(\nu))^2}$,
\end{center}
\vspace{.1in}
as the mass estimator; this minimizes the uncertainty resulting from variability
of the $p_T(W)$ distribution.

\vspace{-.1in}
\paragraph{Gedanken Problem.}
Under what circumstances is lepton transverse energy a better mass estimator
than transverse mass?
\vspace{.1in}

In general, the $Z$ sample can be used to calibrate all responses for
measuring the $W$ mass, but the cross section times leptonic branching
ratio for $Z$s is an order of magnitude smaller than for $W$s.

The measurement depends crucially on calibrating the lepton energy scale.
For the CDF magnetic detector this involves using $\psi \rightarrow \mu\mu$
to calibrate tracking, and understanding dE/dx and tracking systematics
to extrapolate to the higher momenta of $W$ and $Z$ muon decays. The
calibration is transferred to electron measurement by understanding
the tracking material radiation length, and matching predicted E/p for
$W$ electrons. So far, for the most recent large data sample, CDF analysis
is ongoing and only a preliminary muon result has been quoted.\cite{Randy}

\begin{figure}
\vspace{.1in}
\hspace{.5in}\epsfysize3.6in\epsffile{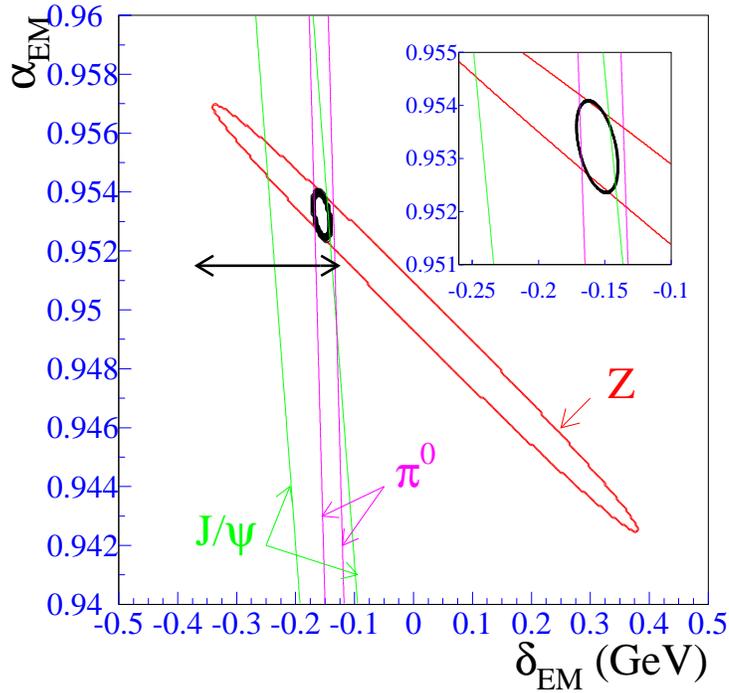}
\caption{The D\O\ electron energy scale determination. $\alpha$ is the scale
and $\delta$ is the effective pedestal offset.}
\end{figure}

The D\O\ detector had no magnet for the data taken so far. They use 
the $Z \rightarrow ee$ mass with
angular correlation, $\psi \rightarrow ee$, the $\pi^0$ mass, and the
test beam measured linearity of their calorimeter to set the electron 
energy scale.\cite{d0mw}
Taking $E(measured) = \alpha_{EM} \times E(true) + \delta_{EM}$, 
the constraints are shown in
Fig. 19. For the final uncertainty, the deviation from linearity allowed
by the test beam data is accounted, expanding the allowed region.

In reconstructing the recoil to $p_T(W)$, the D\O\ calorimeter 
reconstructs more than $80\%$ of the net transverse energy.  By contrast, CDF
reconstructs slightly less than $60\%$. Most of the difference is
due to the absence of a magnet in D\O; D\O\ has a statistical advantage due
to the resulting better transverse mass resolution. The leptonic $Z$ samples
are used to calibrate this response.

\vspace{-.1in}
\paragraph{Gedanken Problem.}
Do you know of any instance when making a compensating
calorimeter (D\O\ as opposed to CDF, ZEUS as opposed to H1), 
as discussed by J. Virdee, created
an actual physics measurement advantage?
\vspace{.1in}

There is no analytic form to describe the transverse mass distribution,
shown for the D\O\ 94-95 data in Fig. 20. A fast Monte Carlo generator including
all that is known about $W$ production and detector response is used
to make transverse mass templates as a function of assumed $W$ mass.
The templates are then used in a likelihood fit. We will see this procedure
of fitting data to Monte Carlo templates several times.

\begin{figure}
\vspace{.1in}
\hspace{.2in}\epsfysize2.0in\epsffile{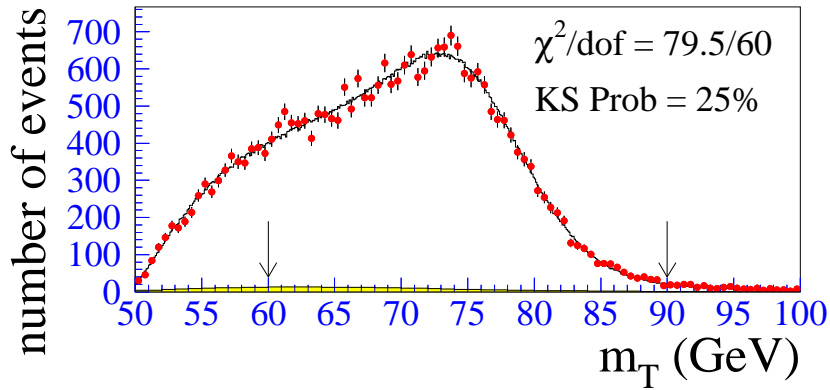}
\caption{The D\O\ electron $W$ transverse mass distribution. The
fitting region is marked, as is the small background contribution.}
\end{figure}

\subsection{RESULTS AND PROSPECTS}

\begin{table}[htb]
\begin{center}
\caption{Measurement errors in MeV/c$^2$ for the 94-95 D\O\ $W$ mass determination.}
\begin{tabular}{lr}
\hline
Statistics & 95 \\
Lepton systematics & 35 \\
Recoil measurement & 40 \\
QCD model (PDF, resum) & 25 \\
QED radiative cor. & 15 \\
Background & 10 \\
Other & 15\\
ALL systematics & 70 \\
\hline
ALL & 120 \\
\hline
\end{tabular}
\end{center}
\end{table}

D\O\ has completed and published their analysis.\cite{d0mw} The errors are
listed in Table 3.  Note that the 95 MeV statistical error includes,
in quadrature, 65 MeV for $Z$ statistics in setting the electron energy
scale. The overall modelling (theory)
error is at the level of $\sim\pm30$ MeV which bodes well for the future.

They measure 
m(W) = 80.44 (12) GeV/c$^2$ for 94-95 data, and 80.43 (11) when combined
with their earlier data.  The preliminary 94-95 CDF muon measurement
gives 80.43 (16), and combined with previous measurements gives
80.38 (12). Combining D\O, CDF, and UA2, and accounting common uncertainties,
gives m(W)$ = 80.41 \pm0.09$ GeV/c$^2$.

D\O\ has reached its goal, while CDF is still working on theirs. Both
detectors are undergoing major upgrades; ``run 2'' 
data taking should begin in 2000.
A solenoid and magnetic tracking is part of the D\O\ upgrade. The accelerator
upgrade includes the Main Injector and much improved antiproton collection.
More bunches will be used to spread out the increased luminosity, the six 
bunches for existing data (``run 1'') will increase
to 36 and eventually to $\sim100$ bunches. 

The run 1 existing datasets correspond to $\sim110$ pb$^{-1}$. By 2003 each 
experiment
hopes to collect 2-4 fb$^{-1}$. Further improvements could produce samples
of perhaps 20 fb$^{-1}$ by the end of 2005. Such samples should allow the
final LEP2 $W$ mass precision of 35-40 MeV/c$^2$ to be 
seriously challenged.

\section{The LEP2 $W$ Mass}

\subsection{GOALS OF THE MEASUREMENT}

The twin goals of the LEP2 program are the study of $W$ bosons and
searches. Part of the former is a goal for an eventual combined
$W$ mass determination to $40$ MeV/c$^2$ or better. My discussion of
progress toward this goal will follow the recent review by 
Glenzinski.\cite{glenz}

\subsection{METHOD OF MEASUREMENT}

With increased RF available, the energy of the LEP ring has risen 
sufficiently to allow $e^+e^- \rightarrow W^+W^-$.  The initial
$W$ mass measurement came from running just over threshold and measuring
the cross section, just as the $\tau$ mass was precisely 
measured at BES.\cite{BES}
The left-most point in Fig. 21 best determines how much the solid curve can
slide horizontally.
This threshold 
analysis is statistically limited, but the drive to search for new things
raised the energy and changed the strategy to direct reconstruction.

\begin{figure}
\vspace{.1in}
\hspace{.6in}\epsfysize3.5in\epsffile{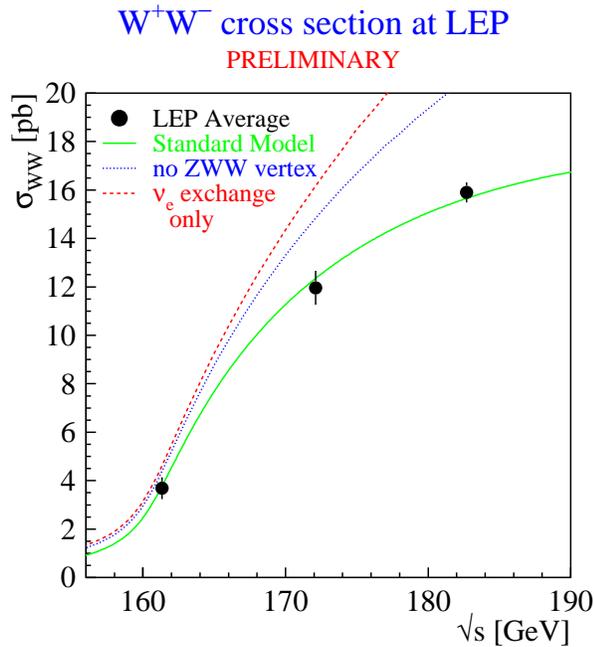}
\caption{The $W$ pair cross section at LEP.}
\end{figure}

The most important channels for direct reconstruction are both $W$s 
decaying hadronically, ``$qqqq$,'' or else 
one decaying hadronically and the other 
producing an electron or muon, ``$\ell\nu qq$.'' Some collaborations include
$\tau\nu qq$, but I will ignore that complication. 
Lepton measurements
and net energy flow have been discussed previously. The identification
of jets of energy flow with final state partons was discussed by Stirling.
The LEP collaborations typically use the JADE\cite{jade} and 
DURHAM\cite{durham} algorithms 
for this analysis.

The analyses use kinematic constrained fits. These apply energy and
momentum conservation to a set of measurements that can be pulled within
their measurement errors, defining a $\chi^2$. For the $qqqq$ case,
energy conservation means the energy of each $W$ is given by the
beam energy, neglecting initial state radiation (ISR). The electron
or positron is likely to radiate a photon down the beampipe before
interacting, calculable in QED.  Each component of momentum should
sum to zero; in the $z$ or beam direction this again implies neglecting ISR.
So one assigns the energy flow to four jets, hypothesizes two pairs as
making up each of the 2 $W$s (there are three choices), 
and calculates a $\chi^2$ for the hypothesis.

In addition to the four constraints (4C) above, one may constrain the two
separate $W$ masses to be the same within an error which includes
the effect of $\Gamma(W)$, 5C.
When good enough solutions are found, either the lowest $\chi^2$ solution,
or a weighted average of acceptable solutions, is used to give the
mass measurement for that event.  The distribution of measurements is fit
to Monte Carlo templates varying the assumed $W$ mass; the Monte Carlo includes
the ISR neglected in fitting. 
Four jet $W$ pair candidates are typically selected with an efficiency of
$\sim85\%$ and a purity of $\sim80\%$. This is illustrated by the 
ALEPH $qqqq$ data at 183 GeV shown in Fig. 22.\cite{ALEPHW}

\begin{figure}
\vspace{.1in}
\hspace{.8in}\epsfysize3.in\epsffile{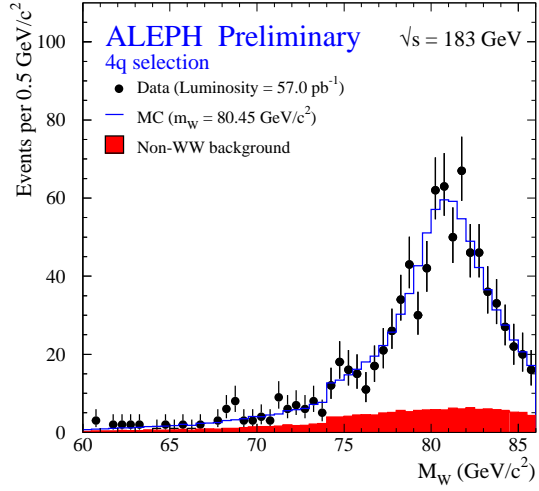}
\caption{The ALEPH $W$ mass reconstruction at 183 GeV in $qqqq$.}
\end{figure}

The event selection for $\ell\nu qq$ is typically $\sim87\%$ efficient and
$95\%$ pure. Since net momentum measurement must be used for the
neutrino, we are left with one constraint, or two for equal $W$ masses.
This channel is illustrated for ALEPH $e\nu qq$ data at 183 GeV in
Fig. 23.\cite{ALEPHW}
Note that in both cases, the use of the beam energy makes the result
less sensitive to absolute jet energy measurement systematics.

\begin{figure}
\vspace{.1in}
\hspace{.8in}\epsfysize3.in\epsffile{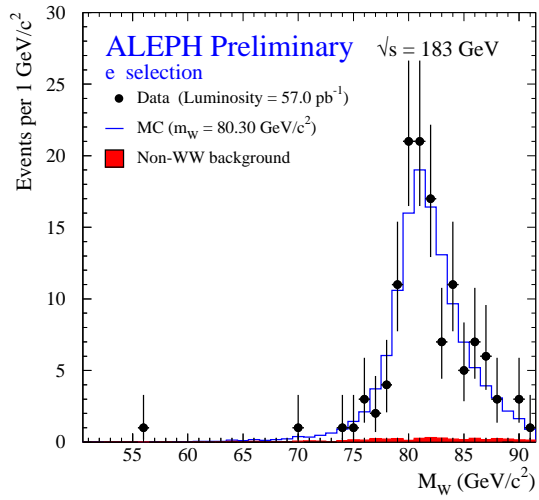}
\caption{The ALEPH $W$ mass reconstruction at 183 GeV in $e\nu qq$.}
\end{figure}
 
\vspace{-.1in}
\paragraph{Gedanken Problem.}
Why is there a tail on the high side in Fig. 23? Does raising the
LEP energy help the $W$ mass measurement by direct reconstruction?
\vspace{.1in}

The DELPHI collaboration uses a somewhat different approach, as discussed
in the talk by Martijn Mulders.  
They attempt to measure ISR event by event, and fit
reconstructed mass to an analytic form.\cite{wdelphi}

LEP runs at the $Z$ from time to time to help monitor detector responses.

\subsection{RESULTS AND PROSPECTS}

\begin{table}[htb]
\begin{center}
\caption{ALEPH 183 GeV $W$ mass systematics 
in MeV/c$^2$. Those marked * are correlated in all experiments.}
\begin{tabular}{lcc}
\hline
SOURCE & $\ell\nu qq$ & $qqqq$ \\
\hline
ISR* & 5 & 10 \\
Hadronize* & 25 & 35 \\
Detector & 22 & 24 \\
Fit & 15 & 14 \\
Beam energy* & 22 & 22 \\
CR/BE* & - & 56 \\
\hline
\end{tabular}
\end{center}
\end{table}

ALEPH preliminary systematic errors for 183 GeV measurements
are given in Table 4.\cite{ALEPHW} ``Hadronize'' refers to 
the systematics of associating
measured energy with final state partons, including soft gluons. 
Detector effects include
boundaries and linearity.  The fit error includes selection biases,
background uncertainties, and Monte Carlo statistics. 
``CR/BE'' refers to QCD final state correlations:
color reconnection, as the $W$s decay so quickly that decay parton color
fields from
both $W$s overlap, and Bose-Einstein correlations for final state
pions from both $W$s. Such final state correlations are being looked for, and
can be measured or limited.

The 183 GeV measurements, quoted at Vancouver for the combined LEP
experiments\cite{LEPEWWG} are m(W) = 80.28 (12) for $\ell\nu qq$ and
80.34 (14) for $qqqq$. The overall direct reconstruction result is
80.36 (9). Combining that with the threshold measurement of 80.40 (20) gives
$m(W) = 80.37 (9)$ for the overall LEP2 direct measurement.
Combining this with hadron collider measurements one obtains\\
\vspace{-.1in}
\begin{center}
$m(W) = 80.39 \pm0.06$ GeV/c$^2$\\
\end{center}
\vspace{.1in}
as the overall direct measurement, with if anything, a shortage of $\chi^2$
in combining results.

The LEP2 run is just getting started. Progress is being made on systematics
so that by the end of the run in 2000, the desired precision looks
possible.

\section{The Tevatron Top Mass}

\subsection{GOALS OF THE MEASUREMENT}

Having established the presence of the top pair signal, the goals
are to confirm the characteristics of that signal in all possible ways
and relevant here, to measure the top mass as accurately as possible.

\subsection{METHOD OF MEASUREMENT}

At the Tevatron, top is predominantly pair produced in the subreaction
$\bar{q}q \rightarrow g \rightarrow \bar{t}t$. Top decays are essentially
$100\%$ to $Wb$, so the final state for top pairs contains two $W$s
like LEP2, a $b$ and $\bar{b}$, and of course, X. Like LEP2, the channels
are characterized by the $W$ decays, again taking leptons as $e$ or $\mu$,
all hadronic ``$qqqqbb$,'' lepton plus jets ``$\ell\nu qqbb$,'' and dilepton 
``$\ell\nu \ell\nu bb$.'' The all hadronic channel has relatively poor 
signal-to-noise.
The dilepton channel has poor statistics, and is underconstrained due to
the two neutrinos, and thus is tricky. I will discuss the lepton plus
jets channel, which gives most of the precision.

\begin{figure}
\vspace{1.3in}
\hspace{.6in}\epsfysize2.4in\epsffile[20 125 550 500]{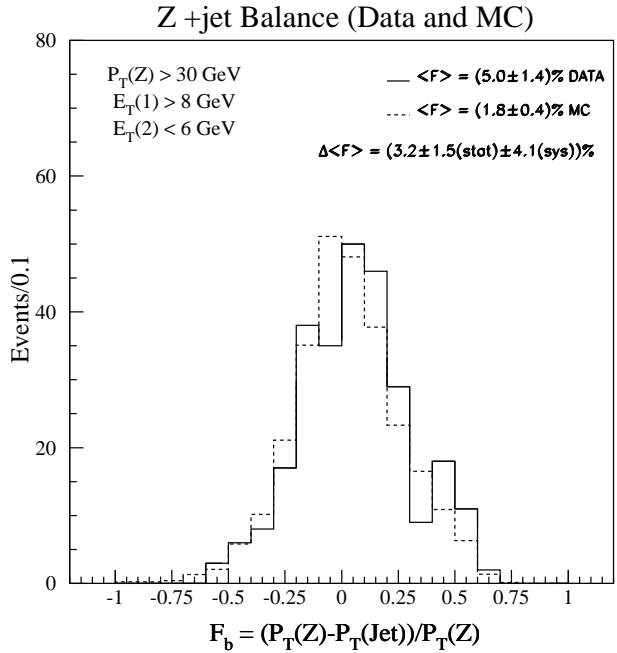}
\caption{The $Z$ jet balance: lepton $p_T$ minus jet $p_T$ over lepton $p_T$
for CDF $Z + jet$ events, data and Monte Carlo.}
\end{figure}

This measurement is based on jet energies, without the beam energy calibration
possible at LEP.  Jets are defined as calorimeter clusters of energy in a
circle (``cone'') in 
$\eta\times\phi$ space of radius typically $\sim0.4$. We have seen, in
the $W$ mass discussion, the EM calorimeter calibration. For CDF, the hadronic 
calibration starts from the test beam, then uses jet fragmentation, with
the nonlinearity of calorimeter response measured from test beam 
and {\it in situ}
isolated particle measurements. There are corrections for final state hadrons
coming from the relevant parton falling outside the cone, and for X getting into
the cone.
As luminosity varies, X varies, which can be corrected on average for a given 
sample; 
there is also a jet threshold bias. These effects are studied by combining top
events, real and Monte Carlo, with one or more minimum bias events.

\begin{figure}
\vspace{.1in}
\hspace{.6in}\epsfysize4.0in\epsffile[70 200 545 700]{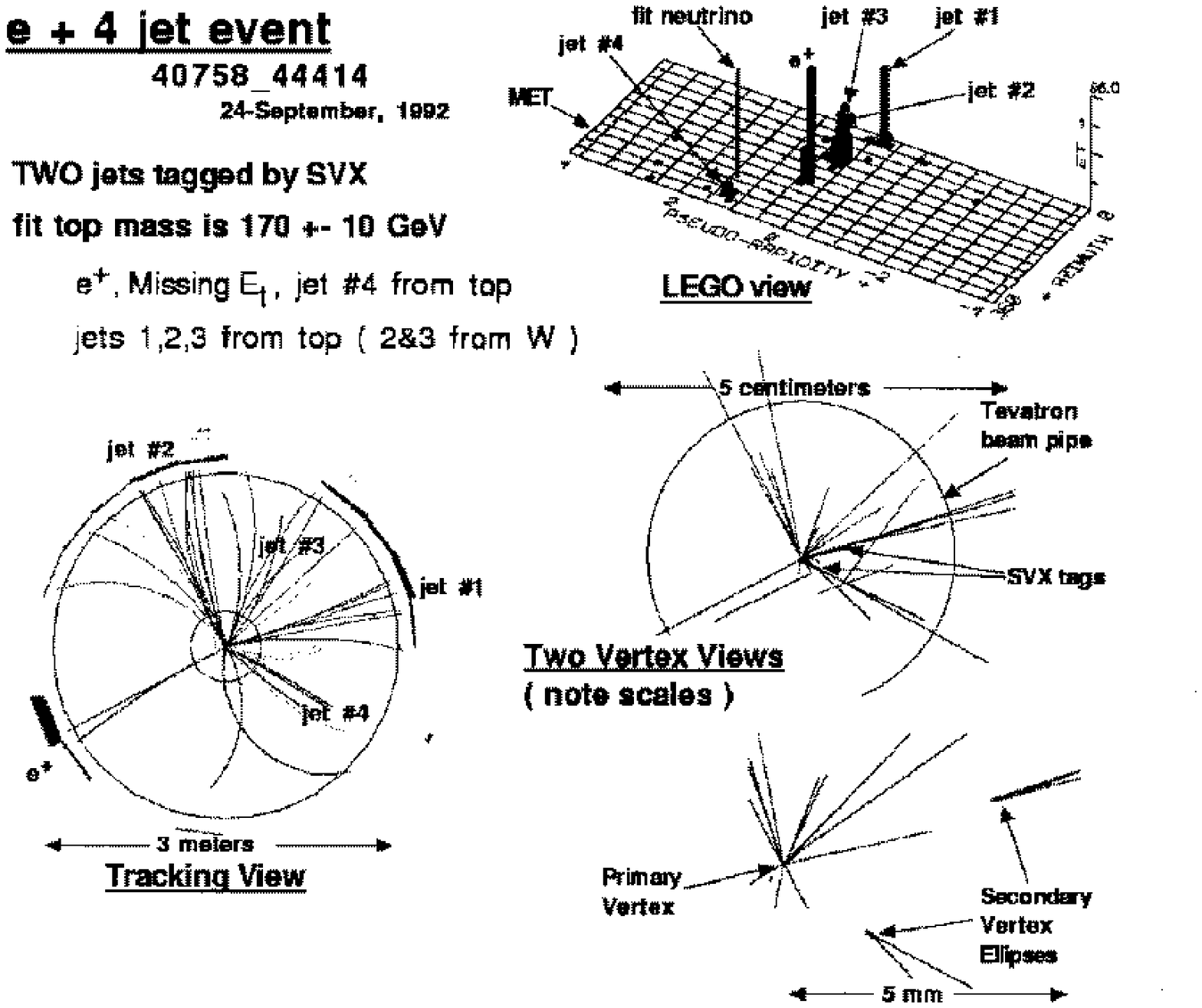}
\caption{A reasonably convincing CDF electron plus four jet top candidate.
The LEGO view shows calorimeter energy in $\eta\times\phi$; the other
views are in $\phi$.}
\end{figure}

The detailed jet response study is done with the help
of tracking, that is for calorimeters covering 
central rapidity. The scale is transferred to the rest
of the calorimeters using dijet balance. The process can be checked using
photon/jet event balance, but one needs to worry that selecting or triggering
on a photon puts an initial state intrinsic transverse momentum 
``$k_T$'' bias. The initial partons tend to
be moving in the photon direction.  
Top specific jet corrections account mostly for muons in $b$ decays.
The final check that this all makes sense is to look at
the transverse energy balance in $Z jet$ events. This is shown in Fig. 24
for CDF $Z$ events, where the lepton pair $p_T$ is above 30 GeV/c. This
is balanced essentially 
by one jet, with no other jet above 6 GeV $E_T$. The agreement is
well within the expected systematic error. 

D\O\ employs an equivalent procedure,
starting from the electromagnetic scale and studying photon jet balance etc.
Although with no magnet, they have fewer handles but their corrections are 
smaller.  

\begin{figure}
\vspace{.1in}
\hspace{.7in}\epsfysize3.5in\epsffile[5 145 540 685]{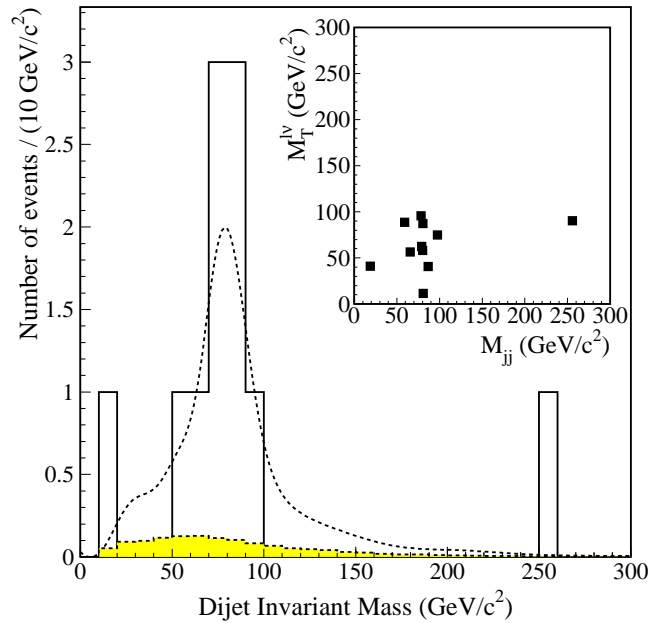}
\caption{The untagged two jet mass for CDF lepton plus 4 jet events where
two jets have 
loose SVX tags, and the correlation of that with the transverse mass
reconstructing the leptonic $W$. The curve is Monte Carlo prediction with
the shaded area background.}
\end{figure}

For obtaining a mass estimate, kinematic fitting is again used. 
One starts with a sample of events with a 
lepton plus four jets and missing $E_T$.
Only overall
transverse momentum balance is available, and that is used to define the neutrino
$p_T$. The longitudinal momentum of the neutrino is unmeasured.
There are two $W$ mass constraints, one on a jet pair and the other on
the lepton and the inferred neutrino. There are usually 
two viable constraint solutions for the neutrino
longitudinal momentum (``ambiguity''). 
A further constraint comes from demanding that the
two top masses be consistent. The net constraint is 2C. 

The fit is well
illustrated by considering an event, shown in Fig. 25, where two
jets are identified as $b$s by silicon vertex tags (``SVX''). 
Jets 2 and 3 seem
together in the calorimeter display, but are clearly separated in the
tracking view.
The secondary vertices at a few mm identify jets 1 and 4 as $b$s.  Thus
jets 2 and 3 should be a $W$; the invariant mass of the pair, not fit, 
is 79 GeV/c$^2$. The only ambiguities are: which $b$ goes with which $W$,
and which neutrino solution to use. The best $\chi^2$ assigns jet 4 to the 
leptonic $W$, and gives an event top mass of $170 \pm10$ GeV/c$^2$. The $W$s
observed for all candidates events,
when there are two loose SVX tags, are shown in Fig. 26. With
enough statistics the $W$ peak may become a calibration.

If you don't have the $b$ tag 
information, you simply try all the possibilities.
Sometimes the best $\chi^2$ is the wrong combination; the top
mass resolution degrades as the number of possible combinations grows. 
Both CDF and D\O\
tag $b$ jets using associated e or $\mu$ from heavy quark decay, called
soft lepton tag, SLT. The statistics, signal and background levels, 
and resolution are
illustrated for CDF 
in Fig. 27. In defense of the SLT sample, it should be noted
that SLT tagged events that are also SVX tags are removed
from the SLT sample and kept as SVX, just as all tagged
events have been removed from the no tag sample.\cite{cdftop}

\begin{figure}
\vspace{.1in}
\hspace{.7in}\epsfysize3.5in\epsffile[15 140 530 685]{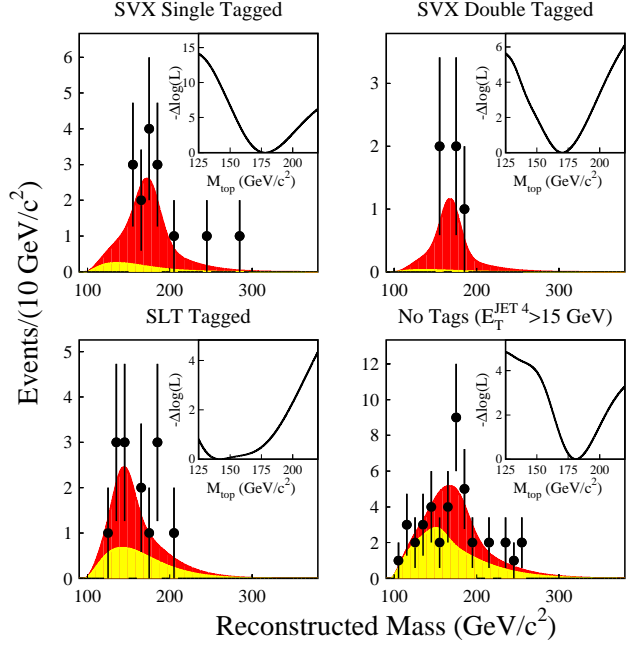}
\caption{CDF top mass distributions and fits for the SVX 1 tag, SVX 2 tag,
SLT tag and no tag samples. The insets are likelihood results. The shaded
areas are the fit results with dark signal and light background.}
\end{figure}

D\O\ defines four variables, with minimal mass bias, that discriminate the top
from the background. The background is 
predominantly $W + jets$ with some fake leptons. 
The variables are missing $E_T$, acoplanarity, the centrality of the
non-leading jets, and a measure of the smallest 2 jet separation.  They do
a joint fit to a discriminant or a neural net (NN) output constructed from
the four variables, and top mass, as
illustrated in Fig. 28. The discriminant and neural net analyses are
combined, accounting correlations, to give their result.\cite{d0top}
The combination uses the technique of pseudoexperiments, analyzing many
Monte Carlo samples of the same size as the data to understand measurements
and correlations. Such exercises allow you to understand whether a
given fit result makes sense.

\begin{figure}
\vspace{.1in}
\hspace{1.in}\epsfysize3.5in\epsffile{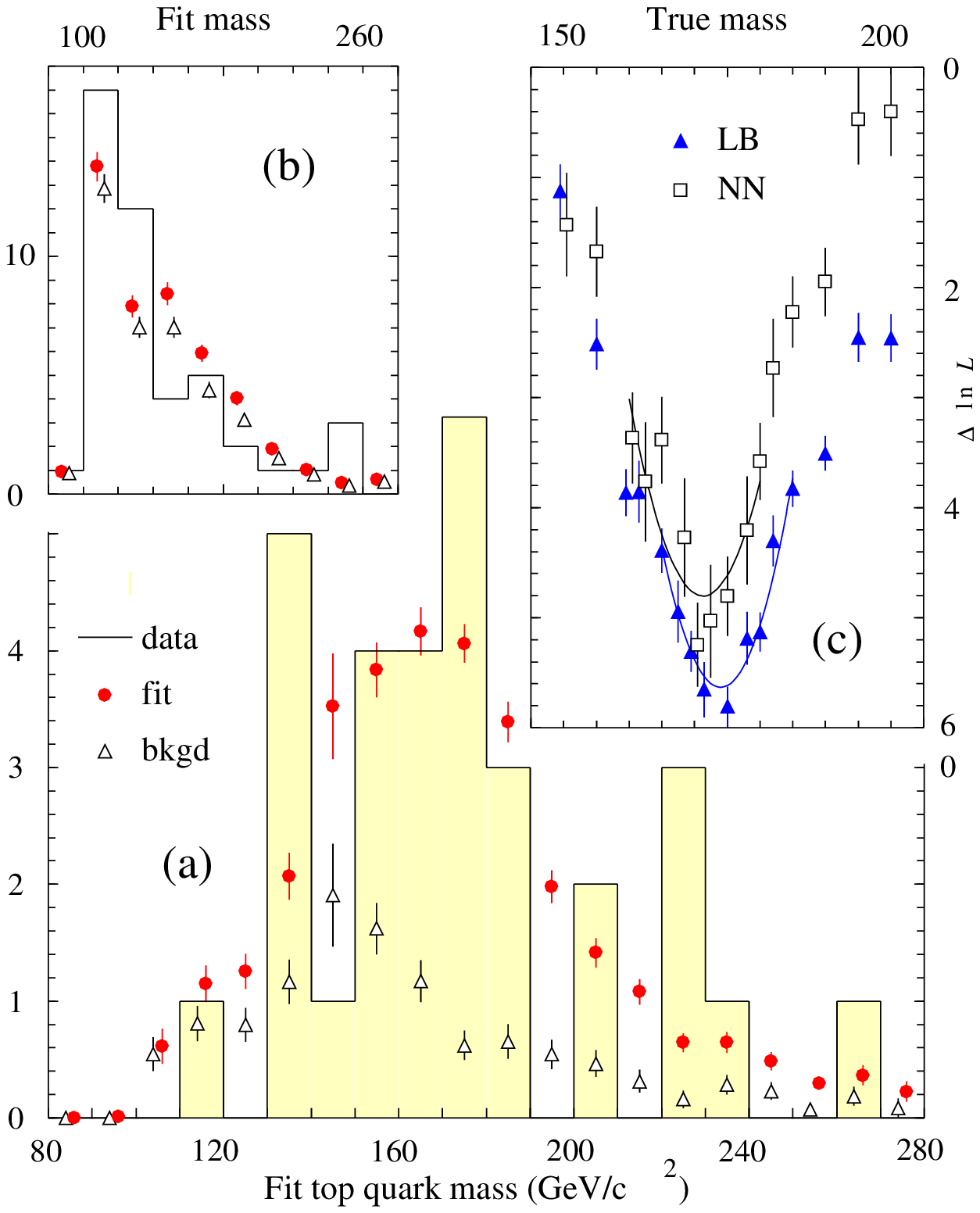}
\caption{D\O\ top mass distribution and fit for (a) predominantly signal
events and (b) predominantly background events. The likelihoods for the
two discriminants are shown in (c).}
\end{figure}

\subsection{RESULTS AND PROSPECTS}

\begin{table}[htb]
\begin{center}
\caption{Top mass measurements, in GeV/c$^2$.}
\begin{tabular}{lcc}
\hline
Channel & Experiment & Value \\
$\ell\nu qqbb$ & CDF & $175.9 \pm6.9$ \\
$\ell\nu qqbb$ & D\O & $173.3 \pm8.4$ \\
$\ell\nu \ell\nu bb$ & D\O & $168.4 \pm 12.8$ \\
$\ell\nu \ell\nu bb$ & CDF & $167.4 \pm11.4$ \\
$qqqqbb$ & CDF & $186 \pm13$ \\
\hline
\end{tabular}
\end{center}
\end{table}

The several top mass determinations are listed in Table 5.
The predominant systematic error comes from the jet energy scale. This
is $\sim\pm5$ GeV/c$^2$ in lepton plus jets.
It includes jet systematics, as discussed, as well as variation with
different assumptions about how much gluon radiation goes where. With
enough data, even the gluon variation can be constrained.
The several results have been combined,\cite{partrid} accounting correlations,
to give\\
\vspace{-.1in}
\begin{center}
$m(top) = 173.8 \pm5.0$ GeV/c$^2$\\
\end{center}
\vspace{.1in}
Only minor refinements to the analysis of existing data may be expected.

It is difficult to predict how much improvement can be achieved with
the Tevatron upgrades, given the jet energy scale systematic level.  
But our measurements with $\sim110$ pb$^{-1}$ are almost as precise as
we predicted
a few years ago for the upgrade 2 fb$^{-1}$
samples.\cite{tev2000} A factor of two improvement
seems reasonably safe, for twenty times the data with improved detectors.

\section{The LEP2 Higgs Search}

\subsection{GOALS OF THE MEASUREMENT}

One of the accomplishments of the LEP program has been
to search for the Standard Model Higgs over the complete kinematically allowed
range of the Higgs mass. So far nothing has been found. In the 
LEP2 era, as the energy
rises the search is extended, typically to twice the beam energy minus the
$Z$ mass. As an illustration, I will describe the L3 analysis of the
183 GeV data.\cite{l3h}

\subsection{METHOD OF MEASUREMENT}

For incrementally adding at the high end of the Higgs search, the relevant
process is $e^+e^- \rightarrow Z^* \rightarrow HZ$,
where $H \rightarrow \bar{b}b$ since
$b$s are the heaviest available decay mode. The final states looked at,
listed as $HZ$, are $bbqq$, $bb\nu \nu$, $bb\ell\ell$, 
$\tau\tau qq$ and $bb\tau \tau$.
Given the four competing experiments, the analysis is fairly sophisticated.
A sample is selected with cuts, a neural net (NN) discriminant is used
to characterize Higgs signal versus background, as in the D0 top mass analysis,
independent of 
mass. A candidate sample is checked as a function of mass and the mass
information is used in defining a purity used to set a limit. The 
various channels are combined for a final result.  

\vspace{-.1in}
\paragraph{Gedanken Problem.}
What are the good and bad points of having several competing
experiments? How many are appropriate?
\vspace{.1in}

For the $bbqq$ mode, the initial sample is
JADE algorithm\cite{jade} 4 jet events. 
Further selection is based
on tracks, calorimeter clusters, visible energy, small net energy flow,
and no lepton or photon candidates. Energy is apportioned to the jets with
the DURHAM algorithm\cite{durham} 
and a 4C fit defines a kinematic $\chi^2_{fit}$.
How well an appropriate jet pair gives a $Z$ mass determines the selection on a
$\chi^2_{mass}$. The 321 events selected compare to 315 predicted for background,
mainly $W$ pairs and $qq\gamma$. For the NN,
tracks, clusters, event shape, $b$ tag and $\chi^2_{fit}$ are used.
The results are shown in Fig. 29. Twenty events have NN$>0.5$, but there
is no sign of any signal.

\begin{figure}
\vspace{.1in}
\hspace{1.in}\epsfysize3.0in\epsffile{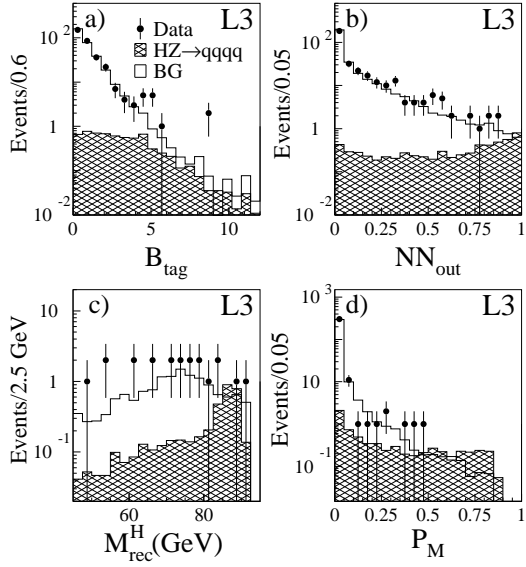}
\caption{For $bbqq$; the $b$ tag discriminant (a), NN output (b), 
$H$ mass for the events with NN$>0.5$
(c) and purity (d). The open histogram
is background, and the shaded one is a nominal 87 GeV Higgs signal.}
\end{figure}

For the $bb\nu \nu$ mode, events are selected on tracks and clusters.
Two jet events (DURHAM) with a recoil mass between 40 and 115 are consistent
with $Z \rightarrow \nu\nu$. Net energy flow and $b$ tag probability
are required. The 56 events found compare to 50 predicted as background.
The NN uses $b$ tag, angles, recoil mass, jet masses, net $E_T$
and $\chi^2_{fit}$. The results are shown in Fig. 30. Again there is no sign
of a signal.

\begin{figure}
\vspace{.1in}
\hspace{1.in}\epsfysize3.0in\epsffile{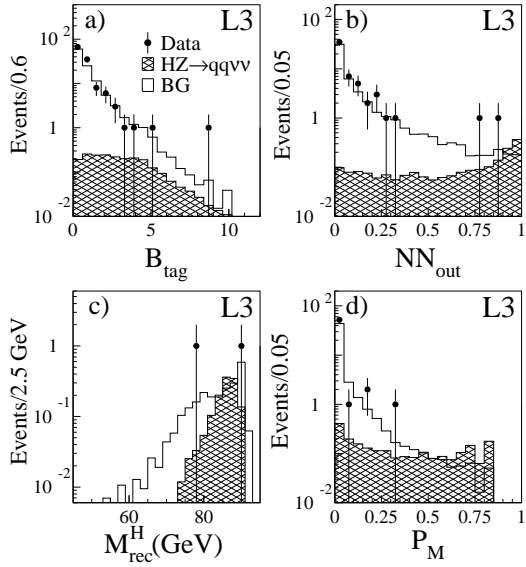}
\caption{For $bb\nu\nu$; the $b$ 
tag discriminant (a), NN output (b), $H$ mass for the two events with NN$>0.5$,
(c) and purity (d). The open histogram
is background, and the shaded one is a nominal 87 GeV Higgs signal.}
\end{figure}

For the $bb\ell\ell$ mode, where $\ell$ 
as usual is $e$ or $\mu$, the $Z$ is observed as the 
lepton pair; 6 $ee$ candidates and 2 $\mu\mu$ candidates are found, again
consistent with background. The NN uses $b$ tag, angles, m(Z), jet masses,
and $\chi^2_{fit}$. Only one event has NN$>0.1$; it gives a mass less than 70.

For the two $\tau\tau$ modes, $\tau$s are selected as two isolated 1/3 prongs
of opposite charge. A cut discriminant uses jets, angles, masses, $b$ tag
and $\chi^2_{fit}$. One event is selected with a predicted background of 2.4.

There is no sign of a signal and systematics are included in a pseudoexperiment
exercise for combining channels. Systematics include luminosity ($0.3\%$),
detector efficiency ($4\%$) and background uncertainty, taken as correlated
between channels ($10\%$).

\subsection{RESULTS AND PROSPECTS}

The L3 analysis sets a limit $m(H) > 87.6$ GeV/c$^2$ at $95\%$ CL. 
The pseudoexperiment
study gives a probability for obtaining a higher limit given the data sample 
of $35\%$, so they did
not get too lucky. The overall LEP2 limit is\cite{Peccei}\\
\vspace{-.1in}
\begin{center}
m($H$) $> 89.8$ GeV/c$^2$ $95\%$ CL.\\
\end{center}
\vspace{.1in}

Considerably more luminosity is expected for LEP2, but more important for
this search, the energy should get up to 200 GeV. Thus, the limit could
get up to 109, or else a signal could be well established up to 99 GeV/c$^2$.

A further window for the Higgs search
to about 120 GeV/c$^2$, will be available at the upgraded Tevatron
if that collider produces data samples of $\sim20$ fb$^{-1}$. There, one
hopes to see $\bar{q}q \rightarrow W^* \rightarrow HW \rightarrow bb\ell\nu$.
It would be interesting to observe $HW$ if $HZ$ is found at LEP, a
possibility which should be kept in mind.  
Beyond that, the possibilities are rather thoroughly covered by LHC, as
discussed by F. Pauss.

\section{Conclusions}

The program of precision electroweak measurements is a great success 
since all the measurements are consistent. The simple-minded minimal
Higgs scenario is still allowed. If the motivation was to get beyond the
limitations of the Standard Model by detecting a contradiction, then
we must report failure.

\subsection{DIGRESSION ON TOPICS NOT COVERED}

Before we get into global fits, a few measurements need to be mentioned.
The decay fraction of $Z$ to $\bar{b}b$, $R_b$, has an interesting history;
the formerly exciting discrepancy is now merely a slight pull toward lower
top mass.  While $Z$ asymmetry measurements dominate the overall Higgs
mass constraint, with $W$ mass in a distant second place, the $Z$ width and
leptonic branching fractions also have some influence.

One can search for new physics in the trilinear couplings, {\it eg.}
$ZWW$. While the destructive interference in $W$ pair production
was demonstrated at the Tevatron,\cite{ct} the 
Tevatron constraints\cite{d0kl} are
being overtaken by the LEP2 studies.\cite{LEPEWWG}

The $W$ branching ratios are now being directly measured at LEP2. The leptonic
fractions will
soon be more precise than the Tevatron constraints from the 
leptonic $Z/W$ cross section 
ratio. The Tevatron still has much larger samples, given
a trigger signature, and continues to have the more accurate direct $W$ width
measurement, and better reach for rare $W$ decays.

I also note that in many cases
flavor universality is tested as well as assumed. 
Axial and vector couplings can be measured separately, and many assumptions
can be relaxed and checked.

\subsection{GLOBAL FITS}

\begin{figure}
\vspace{.1in}
\hspace{.7in}\epsfysize4.5in\epsffile{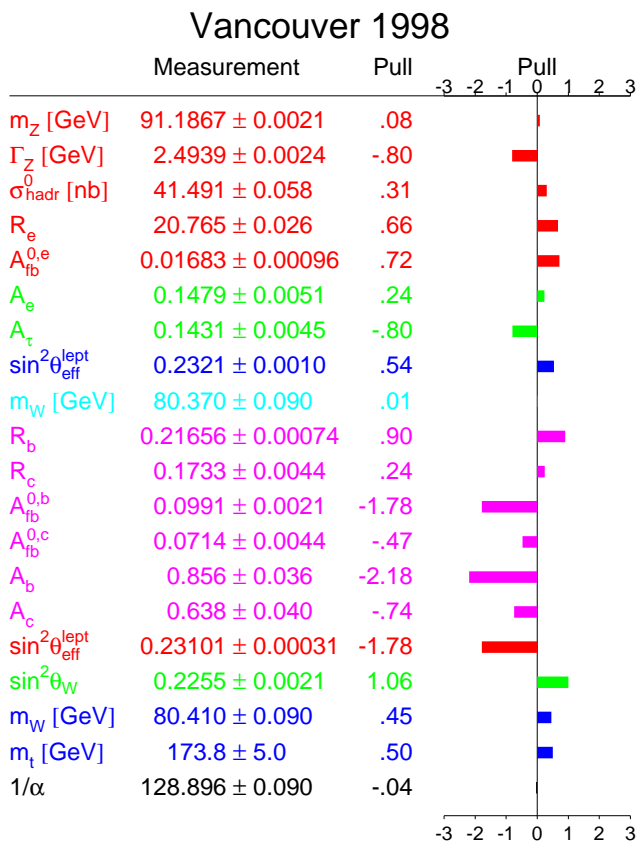}
\caption{The summer 1998 LEPEWWG global fit pull distribution. The top
15 measurements are combined LEP results, then SLD, NuTeV, 
two from the Tevatron Collider, and calculated $\alpha$. The fit
also gives an $\alpha_S$ in agreement with other measurements.}
\end{figure}

Although the g-2 experiment may not soon 
provide input to the global electroweak
fits, all the other measurements discussed 
do. I will use the LEPEWWG version as quoted at Vancouver;\cite{LEPEWWG} 
PDG gets similar results.\cite{PDG} That it all hangs together is shown in
the pulls plot, that is how much each measurement deviates from the fit value,
Fig. 31. The usual suspects are off a bit, but none have
either the statistical significance or the connection to
a popular SUSY scenario to make
them noteworthy.

Let us check if the $W$ mass,
calculated indirectly through radiative corrections, agrees with the direct
measurements.  This is shown in Fig. 32. The direct measurements agree
among themselves, perhaps too well. The NuTeV analysis inputs the measured
top mass, while LEP/SLC does not; they also indirectly infer a top mass.
For the global indirect $W$ mass, the direct top mass measurement
is included, moving the result up.

\begin{figure}
\vspace{.1in}
\hspace{.7in}\epsfysize4.0in\epsffile[80 140 520 670]{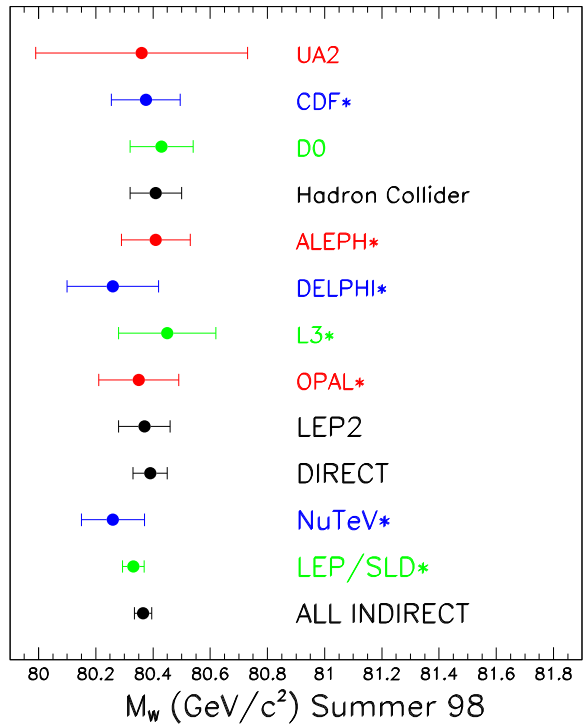}
\caption{The summer 1998 $W$ mass determinations, direct and indirect. 
Measurements marked * include preliminary results.}
\end{figure}

If utility is defined as providing the greatest constraint on the Standard 
Model Higgs mass, then given the incredibly precise measurements of 
$\alpha$, $G_F$ and m($Z$), the $Z$ asymmetries
are most useful.  This constraint is illustrated in Fig. 33. While a
couple of individual measurements would prefer the Higgs to have been 
found some time ago, 
the general trend can accomodate a Higgs mass such that there is
no need for new physics to the Plank scale.

\begin{figure}
\vspace{.1in}
\hspace{1.3in}\epsfysize3.0in\epsffile{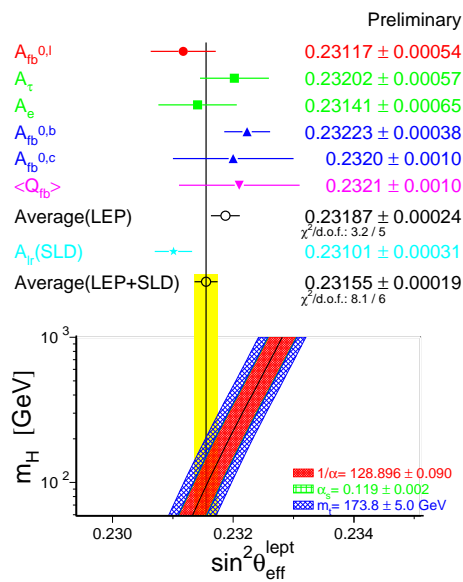}
\caption{The summer 1998 LEPEWWG/SLD $Z$ asymmetry measurements.}
\end{figure}

The correlation of the $W$ mass and the top mass is shown, in the NuTeV
world view, in Fig. 34. The $Z$ asymmetry dominates the width of
the region allowed by LEP/SLC indirect measurements as 
m($H$) changes. A factor of two improvement
on both direct measurements will help a lot.

\begin{figure}
\vspace{.1in}
\hspace{1.in}\epsfysize3.0in\epsffile{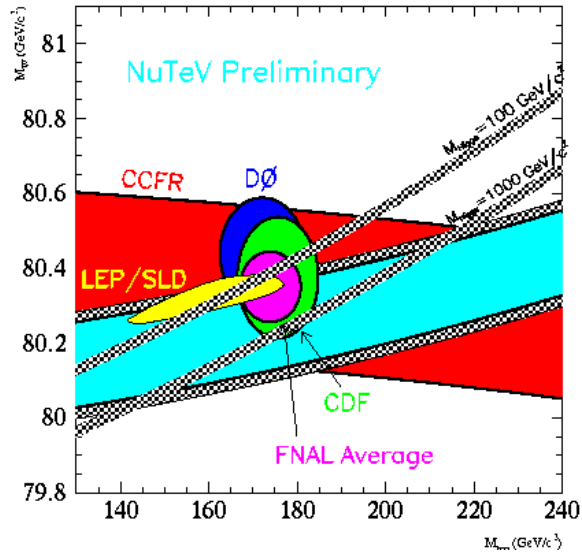}
\caption{The parameter space of $W$ mass and top mass, with bands shown for
Higgs mass values and contours showing measurement constraints.}
\end{figure}

The overall Higgs constraint is shown in Fig. 35. This is quoted as giving a
one-sided $95\%$ CL upper limit on the Higgs mass, increased since
Moriond 1998, of 280 GeV/c$^2$.\cite{KH}  But one should
not really ignore the fact 
that the left side of the plot has been ruled out.  Even in the
most unfavorable MSSM scenario, SUSY Higgs below 70 GeV/c$^2$ 
are ruled out.\cite{l3susyh}
It may be more appropriate to call the limit $\sim90\%$ CL;
$5\%$ of what is left on the right side of the plot 
corresponds to a rather higher mass
limit. No allowance has been made for measurement discrepancies; the
limit depends strongly on the SLD $A_{LR}$ result.  So the SUSY establishment,
hoping that the presence of a low mass Higgs will be established, needs to
remain patient.

The simplest Standard Model Higgs scenario remains viable.
Except for those 
SUSY scenarios which imitate the Standard Model, more complicated
Higgs scenarios generally imply that there is no constraint.

\vspace{-.1in}
\paragraph{Gedanken Problem.}
Using all the information about Fig. 35, and avoiding religious prejudice,
what would you quote for a Higgs mass upper limit?
\vspace{.1in}

\begin{figure}
\vspace{.1in}
\centerline{\epsfysize3.0in\epsffile{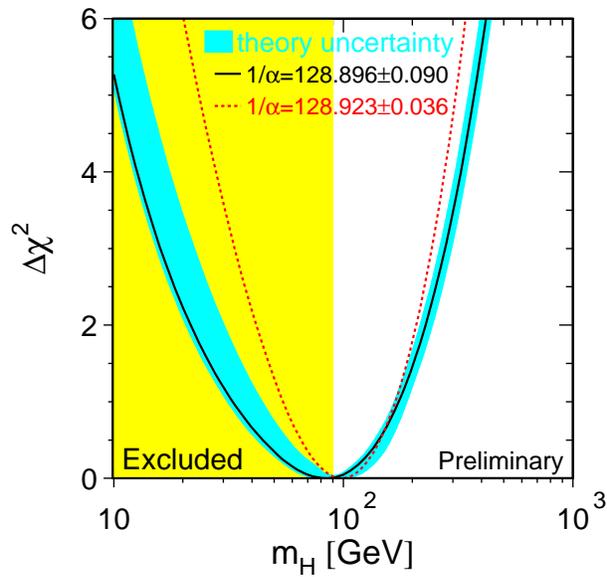}}
\caption{The global electroweak fit $\chi^2$ versus Standard Model
Higgs mass value.}
\end{figure}

\subsection{PROSPECTS}

The BNL g-2 experiment is just getting started. LEP1 and SLC have 
finished running, with some analysis
updates pending. LEP2 is just getting going and one can anticipate
$W$ mass precision improvements, as well as an extension of the Higgs search 
in what seems to be a promising region.  

The Tevatron is on hiatus, upgrading. 
Some updates, particularly on the $W$ mass, are pending.  Once Tevatron running
resumes, substantial improvements may be expected in top mass and $W$ mass
precision, giving these measurements 
comparable electroweak precision to the $Z$ asymmetries.
There is even some window of opportunity to search for
$H \rightarrow \bar{b}b$, slightly extending the LEP2 range.  

If we are persistent and patient, LHC results
must eventually clarify the picture. I certainly hope
that we learn something more than a value for the Standard Model Higgs mass.

\section{Acknowledgements}

I thank Tom Ferbel for being our perfect host. 
I am grateful to the students, and to my fellow lecturers for keeping things
so interesting. I would like to thank Cosmas Zachos, Lee Roberts,
Doug Glenzinski, Jae Yu, Randy Keup, Darien Wood, Alain Blondel and Ruth Hill 
for their help. This work was supported in part
by the U. S. Department of Energy, contract W-31-109-ENG-38.


\end{document}